\documentclass[aip,jcp,preprint,graphicx]{revtex4-1}

\usepackage{graphicx}
\usepackage{amssymb}
\usepackage{amsmath}
\usepackage{color}

\usepackage{overpic}
\newcommand{\cm}{cm$^{-1}$}

\newcommand{\BF}[1]{{\color{black} #1}}
\usepackage{xcolor} 

\let\oldAA\AA
\renewcommand{\AA}{\text{\normalfont\oldAA}}

\begin{document}

\title{On the Role of Non-Diagonal System-Environment Interactions in Bridge-Mediated Electron Transfer}

\author{Nirmalendu Acharyya}
\affiliation{Max-Born-Institut f\"ur Nichtlineare Optik und Kurzzeitspektroskopie, D-12489 Berlin, Germany}

\author{Roman Ovcharenko}
\affiliation{Max-Born-Institut f\"ur Nichtlineare Optik und Kurzzeitspektroskopie, D-12489 Berlin, Germany}

\author{Benjamin P. Fingerhut}
\email[]{fingerhut@mbi-berlin.de}
\affiliation{Max-Born-Institut f\"ur Nichtlineare Optik und Kurzzeitspektroskopie, D-12489 Berlin, Germany}

\date{\today}

\begin{abstract}
Bridge-mediated electron transfer (ET) between a  donor and an acceptor  is prototypical for the description of 
numerous most important  ET scenarios. 
While multi-step ET and the interplay of sequential and direct superexchange transfer pathways in the donor-bridge-acceptor (D-B-A) model is increasingly understood,  
the influence off-diagonal system-bath interactions on the transfer dynamics is less explored.
Off-diagonal interactions account for  the dependence of the ET  coupling elements on nuclear coordinates  (non-Condon effects) and are typically neglected.
 Here we numerically investigate with quasi-adiabatic propagator path integral (QUAPI) simulations the impact of off-diagonal system-environment interactions  on the transfer dynamics for a wide range of scenarios in the D-B-A model. 
 We demonstrate that off-diagonal system-environment interactions can have profound impact on the bridge-mediated ET dynamics.
 In the considered scenarios the dynamics itself does not allow for 
 a rigorous assignment of the underlying transfer mechanism.
 Further, we demonstrate how  off-diagonal system-environment interaction mediates anomalous localization by preventing 
 long-time depopulation of the bridge B and how coherent transfer dynamics between donor D and acceptor A can be facilitated.
 The arising non-exponential short-time dynamics and coherent oscillations are interpreted within an equivalent Hamiltonian representation of a primary reaction coordinate model 
 that reveals how the complex vibronic interplay of vibrational and electronic degrees of freedom underlying the non-Condon effects can impose donor-to-acceptor coherence transfer on short timescales.

\end{abstract}

\pacs{}

\maketitle 

\section{Introduction}
\begin{figure}[b!]
\includegraphics[width=0.48\textwidth]{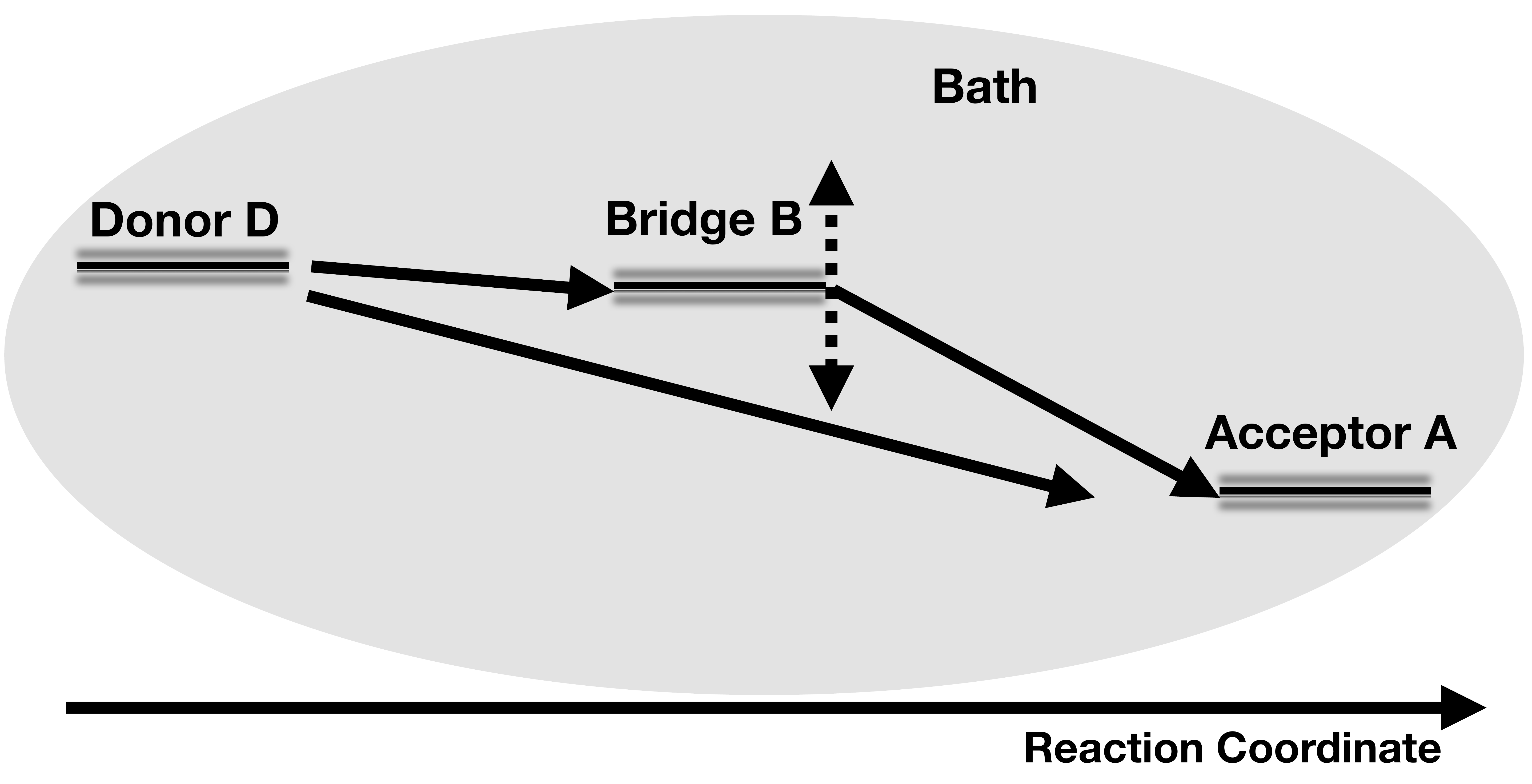}
\caption{{\bf Schematic of the Donor-Bridge-Acceptor (D-B-A) Model.}
The variation of the bridge energy $\epsilon_B$ allows to tune between the sequential D-B-A model with a low energy bridge state ($\epsilon_D > \epsilon_B$) and the D-B-A model with asymmetric-$\Lambda$ configuration ($\epsilon_D < \epsilon_B$).  
Possible transfer pathways are  the sequential $|D\rangle \to  |B\rangle \to  |A\rangle$ and the direct superexchange pathway  $|D\rangle \to  |A\rangle$. 
Diagonal and off-diagonal system-bath interactions lead to fluctuations of state energies $\epsilon_i$ and electron transfer coupling elements $V_{ij}$ that mediate the dynamics.
}
\label{fig:Sc1}
\end{figure}

 Photosynthetic solar
energy conversion relying on molecular charge carriers starts with 
the light-induced generation of Frenkel-type excitons, 
followed by the  irreversible fixation of excitonic energy  in  
multi-step electron transfer (ET) reactions.
In 
reaction centres (RC),
of purple bacteria or plants,  the irreversible fixation proceeds across a phospholipid  membrane in a sequence of  directional and highly efficient ET reactions,
mediated by spatially well organized (bacterio-) chlorophyll and pheophytin molecules. 
Pioneering 
electrostatic considerations\cite{Warshel:PNAS:1981},
 supported by multi-objective evolutionary algorithm optimizations \cite{Fingerhut:PCCP:2010}  suggest that efficient and irreversible charge separation
 requires a sequence of at least three molecular charge carriers.
The emerging prototypical setup of bridge $B$ mediated 
multi-step or superexchange 
ET between a donor $D$ and an acceptor $A$
(see Fig.~\ref{fig:Sc1}), thus serves as minimal model for the description of charge separation in RC 
and has shown 
relevance for numerous bridge-mediated 
ET processes \cite{Bixon:JCP:1997,Nitzan:AnnRevPhysChem:2001,Goldsmith:JPCB:2006,Paulson:JACS:2005,Berkelbach:JCP:2013,Guo:JACS:2019}.

%
%
While nonadiabatic ET theory provides a valuable starting point for the descriptions of multi-step ET reactions, 
e.g. in the ET kinetics of the bacterial RC \cite{Arlt:PNAS:1993,Bixon:CPL:1989,Huppmann:JPCA:2003,Fingerhut:CPL:2008}, limitations arise from the inherent assumptions of instantaneous medium relaxation, 
the perturbative nature of Fermi golden rule expressions and the typically neglected dependence of electronic coupling on nuclear coordinates (Condon approximation).
%
Non-exponential short-time dynamics and coherent oscillations observed in pump-probe 
and, more recently, two-dimensional electronic spectroscopy experiments of bacterial RC suggest a more complex picture 
where details of the ET reaction are  potentially affected by the interplay of coherent electronic and 
nuclear motion.\cite{Vos:Nature:1993,Sporlein:JPCB:1998,Novoderezhkin:JPCB:2004,Ma:NatCom:2019}

Particular theoretical challenges arise for a strong interaction of vibrational modes and  electronic degrees of freedom that can 
impose oscillatory dynamics in bridge mediated ET.\cite{Novoderezhkin:JPCB:2004}
Tanimura investigated 
multistate ET for a system coupled to 
a heat bath with a non-Ohmic spectral density 
with the numerically exact reduced hierarchy equations of motion (HEOM) approach\cite{Tanaka:JPJap:2009,Tanaka:JCP:2010}
%
%
\BF{t}hat is numerically efficient for Debye spectral densities.
Nevertheless, the strong coupling to the environment as typically realized in ET reactions is a persistent challenge for convergence\cite{Kramer:JPCB:2017,Lambert:NatCom:2019}. 
Pioneering quasi-adiabatic propagator path integral  (QUAPI)  simulations by Makri and coworkers provided valuable insight into details of ET in bacterial RC.\cite{Makri:PNAS:1996,Sim:JPCB:1997} 
The QUAPI method formally does not assume
a particular form of the environment spectral density 
or the interaction strength but
 the long system bath memory times of a sluggish environment pose persistent challenges to QUAPI methods.\cite{Richter:2017,Strathearn:2018,Makri:JCP:2020}

Recently, we investigated bridge-mediated ET in a model of the bacterial RC, where charge separation is initiated from  a non-equilibrium excitonic superposition.\cite{Richter:FaradayDisc:2019} 
The path integral simulations 
with particularly longtime system-environment correlations allowed to explore the influence of discrete vibrational modes on the transfer dynamics. While excitonic energy transfer was found to be strongly affected, the 
kinetics of ET  dynamics 
appeared exceptionally robust to the details of the spectral density function, suggesting a picture 
where intramolecular vibrations assure the robustness of optimal, non-activated ET reactions. 


Hence, multi-step ET in the D-B-A model is increasingly understood beyond perturbative approaches and for various 
(non-equilibrium) conditions of the environment.
Nevertheless, the impact of off-diagonal system-bath interactions on the transfer dynamics is less explored.  Such off-diagonal system-environment interactions
 are associated with non-Condon effects, i.e., the dependence of the coupling matrix element mediating the transfer dynamics on the nuclear coordinates.  
 The nuclear coordinate dependence of electronic coupling  was explored in the context of nonadiabatic transitions, indicating that  promoting modes can dominate the vibrational effects.\cite{Goldstein:JPC:1993}
Milischuk and  Matyushov explored non-Condon effects for nonadiabatic electron transfer reactions in donor--bridge--acceptor systems.\cite{Milischuk:JCP:2003}
The importance of non-Condon effects was further  highlighted for ET at oligothiophene-fullerene interfaces 
via multi-layer MCTDH simulations \cite{Tamura:JCP:2012} and the role of  off-diagonal couplings  was emphasized for the formation of charge-transfer states in polymeric solar cells.\cite{Yao:2015aa}
Condensed phase ET beyond the Condon approximation was recently explored by Mavros and  Van Voorhis.\cite{Mavros:2016}
%
\BF{Off-diagonal environment fluctuations were further identified to induce unexpectedly fast Förster resonant energy transfer between orthogonal oriented photoexcited molecules.\cite{Nalbach:PRL:2012}}
A review highlighting limitation of the Condon approximation in biological and bioinspired ET reactions is given in Ref.~\citenum{Skourtis:AnnRevPhysChem:2010}.

%

Here, we show via non-perturbative
 QUAPI simulations that non-diagonal system-environment interactions can have profound impact on bridge-mediated ET  dynamics. 
We investigate the dynamics in different regimes of the prototypical Donor-Bridge-Acceptor (D-B-A) model and demonstrate, that induced by non-Condon effects, the dynamics itself precludes a rigorous assignment of the underlying transfer mechanism. Further, we demonstrate how anomalous localization, mediated by off-diagonal system-environment interaction  prevents the depopulation of  the bridge $B$ and how
coherent transfer between donor $D$ and acceptor $A$ can be facilitated in presence of off-diagonal system-environment interactions.


\section{Theoretical Methods}

\subsection{Multi-Step Electron Transfer with Non-Diagonal System-Environment Interaction.}

We consider a three-state D-B-A model (see schematic in Fig.~\ref{fig:Sc1}) that interacts bi-linearly via diagonal and off-diagonal interactions with the environment:
\begin{eqnarray}
H &=& \left( \begin{array}{ccc}
\epsilon_D & V_{DB}&0 \\
V_{DB} &\epsilon_B &V_{BA} \\
0&V_{BA} & \epsilon_A
\end{array}\right)+ \sum_j\left( \frac{p_j^2}{2 m_j} + m_j \omega_j^2 x_j^2\right) + \nonumber \\
&& \left( \begin{array}{ccc}
d_1  & C_{12}& 0 \\
C_{12} & d_2 & C_{23} \\
0& C_{23} & d_3
\end{array}\right)\sum_j c_j x_j~.
\label{eq:Ham2}
\end{eqnarray}
%
%
%
%
In eq.~\ref{eq:Ham2}, $\epsilon_i$ denote site energies of donor $D$, bridge $B$ and acceptor $A$, and $V_{ij}$ are the respective electron transfer coupling elements. 
The bath is composed of harmonic oscillators with momenta $p_j$, position $x_j$, mass $m_j$ and frequency $\omega_j$. The linear coupling constant $c_j$ mediates the interaction between system and the environment where $d_i$ specify the diagonal system-bath interaction inducing fluctuation of energy levels $\epsilon_i$ and $C_{ij}$ represent off-diagonal system bath interactions that induce fluctuations of the transfer coupling elements $V_{ij}$.
We restrict the Hamiltonian (eq.~\ref{eq:Ham2}) to the case with  $V_{DA}$ = 0, i.e., for diagonal system-environment interaction only the sequential pathway contributes to the dynamics.

We consider the coupling to a single bath affecting site energies $\epsilon_i$ and transfer couplings $V_{ij}$.
For linear coupling all information about the environment is contained in  the spectral density. 
\begin{equation}
J(\omega) = \sum_j \frac{c_j^2}{m_j \omega_j} \delta( \omega -\omega_j). 
\end{equation}
For Ohmic dissipation  
\begin{eqnarray}
J(\omega) =\frac{  \alpha \pi}{2} \omega e^{-\omega/\omega_c}~,
\label{Ohmic_SD}
\end{eqnarray}
$\alpha$ characterizes the system-bath interaction strength and $\omega_c$ specifies the cut-off frequency that is related to the inverse Drude memory time $\tau_D = 1/\omega_c$.
A key quantity in the description of ET  is the reorganization energy $\lambda_R$, defined as $\lambda_R = \frac{1}{\pi} \int \frac{J(\omega)}{\omega} d \omega \approx \alpha \omega_c/2.$


\subsection{Primary Reaction Coordinate Model.}
\label{sec:prc}
The Hamiltonian  with bi-linear diagonal and off-diagonal system-environment interactions  (eq.~\ref{eq:Ham2})  can be equivalently mapped onto a model where the  electronic system  interacts with a primary reaction coordinate $Q$, which couples and dissipates into a bath\cite{Garg:JCP:1985,Lambert:NatCom:2019,Correa:JCP:2019}
\begin{eqnarray}
H &=& \left( \begin{array}{ccc}
\epsilon_D & V_{DB}&0 \\
V_{DB} &\epsilon_B &V_{BA} \\
0&V_{BA} & \epsilon_A
\end{array}\right)  + \nonumber  \\
&&\left( \begin{array}{ccc}
\kappa_1  & \Lambda_{12}& 0 \\
\Lambda_{12} & \kappa_2 & \Lambda_{23} \\
0& \Lambda_{23} & \kappa_3
\end{array}\right)\sqrt{\Omega} Q \nonumber 
 + \frac{P^2}{2} + \frac{ \Omega^2 }{2}Q^2 + \\
 && \frac{1}{2}\sum_j\Big( \frac{p_j^2}{ m_j} + m_j \omega_j^2 \Big(q_j + \frac{\widehat{c}_i Q}{m_i \omega_i^2}  \Big)^2 \Big).
\label{eq:HamPRC}
\end{eqnarray}
Here, $\Omega$ is the frequency of the primary mode and $P$ is the conjugate momenta. 
The representation of the original Hamiltonian with diagonal and off-diagonal interactions in the form of the primary reaction coordinate model (eq.~\ref{eq:HamPRC}) directly reveals the coordinate dependence of  transfer coupling elements $V_{ij}$, i.e., non-Condon effects.
%

The information about the interaction of the primary mode with the environment is contained in the modified spectral density
\begin{equation}
\widehat{J} (\omega) =  \sum_{i} \frac{\widehat{c}_{i}^2}{m_{i} \omega_{i}} \delta( \omega -\omega_{i}). 
\end{equation}
When the relaxation dynamics of the bath is fast, the low frequency part  of $\widehat{J} (\omega)$ becomes relevant which yields an Ohmic spectral density
\begin{eqnarray}
\widehat{J} (\omega) = {\gamma} \omega, \label{Ohmic_SD_PRC}
\end{eqnarray}
where $\gamma$ is the damping coefficient of the primary mode.  

 The primary mode can be expressed as a linear combination of the bath normal modes $\{ x_{j}\}$:\cite{Chernyak:1996}
\begin{eqnarray}
Q = \sum_j v_{j} x_{j}. \label{cannonical_trans}
\end{eqnarray}
Using this canonical transformation (eq.~\ref{cannonical_trans}), the Hamiltonian of the primary reaction coordinate model (eq.~\ref{eq:HamPRC}) can be recast in the form of eq.~\ref{eq:Ham2}
where 
\begin{eqnarray}
&&d_1\,\, c_j = \sqrt{\Omega}\kappa_1 v_j,\quad d_2\,\, c_j = \sqrt{\Omega}\kappa_2 v_j,\quad d_3\,\, c_j = \sqrt{\Omega}\kappa_3 v_j, \nonumber \\
&&C_{12} \, c_j = \sqrt{\Omega}\Lambda_{12} v_j, \quad C_{23} \, c_j = \sqrt{\Omega}\Lambda_{23} v_j.
\end{eqnarray}

Let us denote the eigenvalues of the interaction matrix  (second term in eq.~\ref{eq:HamPRC}) as $\{e_1, e_2, e_3\}$. 
In the overdamped limit, ${\gamma} \gg \Omega$, the parameters of the spectral density $J(\omega)$ (eq.~\ref {Ohmic_SD}) are determined by 
\begin{eqnarray}
\omega_c \approx \frac{\Omega^2}{ {\gamma}},\quad\quad \alpha \approx \frac{ e_1 {\gamma}}{ D_1 \Omega^3}, 
\end{eqnarray}
where $D_1$ denotes the lowest eigenvalue of the original system-bath interaction matrix (see below, eq.~\ref{eq:D}).
On the other hand, a diagonal shift $d_1 \to d_1 +d_0$, $d_2 \to d_2 +d_0$ and $d_3 \to d_3+d_0$ (with $d_0$ being a  real constant)
induces a shift of  the eigenvalues.
In the reaction coordinate model, this corresponds to a 
shifting of the origin of the primary oscillator 
and the displacement of the initial wavepacket via $\kappa_i$.
Thus  $d_0 \neq 0$ implies a displacement of an initial  wavepacket from the Frank-Condon region $Q=0$ which can be used 
 used to reveal non-Condon effects in the dynamics (cf. Sec.~\ref{sec:Coh}).

\subsection{Quasi Adiabatic Propagator Path Integral Simulations with Non-Diagonal System-Environment Interaction}
%
We are interested in the time evolution of the  
 reduced density matrix 
 \begin{equation}
 \widetilde{\rho}(t) = Tr_B \Big[ e^{-i Ht} \rho(0) e^{i Ht}\Big]~,
  \end{equation}
  with $H$ given by eq.~\ref{eq:Ham2},
  which determines observables.
 \BF{F}or diagonal interaction with the environment ($C_{ij}$ = 0) numerical exact simulations are facilitated with the QUAPI
 method\BF{.}\cite{Makri:JCP:1995,Sim:CPC:1997,Sim:JCP:2001}
 For the general interaction matrix 
 \begin{eqnarray}
M \equiv \left( \begin{array}{ccc}
d_1  & C_{12}& 0 \\
C_{12} & d_2 & C_{23} \\
0& C_{23} & d_3
\end{array}\right),
\label{eq:SBM}
\end{eqnarray}
 which satisfy
\begin{eqnarray}
M v_1 =D_1 v_1, \quad M v_2 =D_2 v_2, \quad M v_3 =D_2 v_3, 
\end{eqnarray}
we define the unitary matrix $U$ \cite{}
 \begin{eqnarray}
 U= [ v_1, v_2, v_3]
 \end{eqnarray}
 which diagonalizes 
 the system-bath interaction  $M$:\cite{Lu:JCP:2013}
\begin{eqnarray}
U^\dagger M U =\left( \begin{array}{ccc}
D_1  & 0& 0 \\
0 & D_2 & 0 \\
0& 0 & D_3
\end{array}\right)~.
\label{eq:D}
\end{eqnarray}
Upon transformation of the total Hamiltonian (eq. \ref{eq:Ham2}), we obtain
{\small \begin{eqnarray}
H' &=& U^\dagger H U
\label{eq.9}\\
&=& U^\dagger \left( \begin{array}{ccc}
\epsilon_D & V_{DB}&0 \\
V_{DB} &\epsilon_B &V_{BA} \\
0&V_{BA} & \epsilon_A
\end{array}\right)U+ \sum_j\left( \frac{p_j^2}{2 m_j} + m_j \omega_j^2 x_j\right)+ \nonumber \\
&& \left( \begin{array}{ccc}
D_1  & 0& 0 \\
0 & D_2 & 0 \\
0& 0 & D_3
\end{array}\right)\sum_j c_j x_j ~.
\label{eq.10}
\end{eqnarray}}
The assumption of a single spectral density affecting diagonal and off-diagonal elements of the Hamiltonian (eq.~\ref{eq:Ham2}) facilitates accurate  QUAPI simulations for diagonal and off-diagonal system-bath interactions via unitary transformation. This situation corresponds to the physical situation of an avoided crossing along a  primary reaction coordinate with fluctuating barrier. 
\BF{In the current approach, the fluctuations in the site energies $\epsilon_i$ and transfer couplings $V_{ij}$ are correlated and the results depend on the relative sign of off-diagonal environment interactions  $C_{ij}$. Correlation of fluctuations of site energies and transfer couplings was suggested in simulations of the FMO light-harvesting complex \cite{Olbrich:JPCB:2011} and can introduce  interesting transport phenomena as suggested in the context of a rocking ratchet driven by a single periodic force.\cite{Nalbach:PhysRevE:2017}}
%
%
 The more general case with two spectral densities acting on  diagonal and off-diagonal elements of the Hamiltonian (eq.~\ref{eq:Ham2})\BF{\cite{Krempl:ChemPhys:1996}} was considered in Ref.~\citenum{Mavros:2016} and \BF{recently demonstrated for path integral approaches,\cite{Palm:JCP:2018} it} enables the emergence of conical intersections\cite{Domcke:Book}. 
Note that the unitary transformation (eq.~\ref{eq:D}) is determined by details of the  system-environment interaction and does not necessarily diagonalize the system Hamiltonian. The introduced basis rotation is thus distinct from  transformations solely determined by the system Hamiltonian.\cite{Romero-Rochin:PhysicaA:1989}

Employing factorized initial conditions $\widetilde{\rho}(0)=|D\rangle \langle D|$
 and assuming the bath in thermal equilibrium  at temperature $T$, 
$\widetilde{\rho}(t)$ was evaluated numerically by transforming 
the initial conditions
\begin{eqnarray}
\rho'(t=0) = U^\dagger \rho(t=0) U.
\label{eq:init}
\end{eqnarray}
followed by solving for $\rho'(t)$ in transformed basis using QUAPI methods, and followed by 
reverse transform to obtain $\widetilde{\rho}(t)$.
In particular, numerical propagation was performed with the recently introduced mask
  assisted coarse graining of influence coefficients (MACGIC)-QUAPI~\cite{Richter:2017} method that facilitates the numerical treatment of long-time non-Markovian system-bath correlations while retaining convergence to numerical exact results.
  The algorithm  uses a coarse grained representation of the influence functional (represented by mask of size $k_{eff}$)  to select the relevant paths for propagation for a finite  non-Markovian memory time $\tau_M \propto \tau_D$. Convergence to  numerically exact results  is obtained  by decreasing the size of the Trotter  time-step $\Delta t$ and increasing memory time $\tau_M = \Delta k_{max} \Delta t$ together with 
  increase in the  number of coarse grained quadrature points ($k_{eff} \rightarrow \Delta k_{max}$), details of the  MACGIC-QUAPI~are described in Refs.~\citenum{Richter:2017,Richter:FaradayDisc:2019}\BF{.}
 %
 
 \BF{In the numerical simulations, the employed Ohmic spectral density with exponential high-frequency cut-off (eq.~\ref{Ohmic_SD}) accounts for non-Markovian system-bath correlations on a timescale $\tau_M \approx 100 - 120$~fs ($\Delta k_{max} = 32-64$). Convergence is typically achieved with $k_{eff} \approx 12$ demonstrating the  numerical efficiency of the MACGIC-QUAPI~\cite{Richter:2017} method
 ~(see  Supplementary Material (SM), Table~S.3 and Fig.~S.4 
  for numerical  demonstrations of convergence). Particular challenges in the numerical treatment arise from the unitary transformation (eq.~\ref{eq.10}), where substantially large interstate electronic couplings can arise in intermediate transformed basis for suitable rotation angles imposed by the system environment interaction matrix.
 The  substantial interstate couplings impose the necessity of small propagation timesteps $\Delta t$ for which longtime system-bath correlations are challenging for convergence (e.g. for typical bridge - acceptor energy gaps of 1000~\cm (see below), $V_{BA}^\prime$ can reach 1000 \cm~thereby inducing ultrafast coherence transfer dynamics).
  %
Typical simulations with diagonal system-environment interaction  use filter thresholds $\theta = 10^{-7} - 10^{-8}$ (Table~S.3), account for $\approx 10^4-10^5$ paths during propagation  and show typical run times of a few minutes. For off-diagonal system-environment interaction the number of considered paths increases to $\approx 10^5 -  10^6$ with run times on the order of hours. Convergence tests  employed stricter filter thresholds ($\theta = 10^{-10} $) and accounted for $\approx 10^7$  paths, requiring 12 h (20 cores).
%
%
%
   Notable exceptions are the numerical simulations of anomalous bridge localization (Sec.~
\ref{sec:Loc}) and the demonstration of direct donor-acceptor coherence transfer (Sec.~\ref{sec:Coh}) that require to consider ultralong  memory times up to $\approx$1 ps and 320 fs (SM, Table~S.3), respectively, where in the case of direct donor-acceptor coherence transfer the memory time has to cover the entire  non-Markovian oscillatory dynamics.}

\vspace{1cm}
\section{Results and Discussion}

In the following we investigate the impact of non-diagonal system-bath interactions on the prototypic dynamics of multi-step electron transfer in the bridge-mediated three level system (see Fig.~\ref{fig:Sc1}). We start by investigating the  regime of sequential, bridge-mediated transfer dynamics (Sec.~\ref{sec:seq}) and further examine how
off-diagonal system-bath interactions can activate the direct 
superexchange transfer pathway $|D\rangle \to  |A\rangle$  (Sec.~\ref{sec:SuEx}).
Section~\ref{sec:Loc} presents scenarios how off-diagonal system bath interactions can induce anomalous population  localization in the bridge state $B$ and Section~\ref{sec:Coh} demonstrates  off-diagonal mediated coherence transfer between donor and acceptor states.


\subsection{Sequential Donor-Bridge-Acceptor (D-B-A) Model}
\label{sec:seq}
\begin{figure*}[t!]
\includegraphics[width=1.0\textwidth]{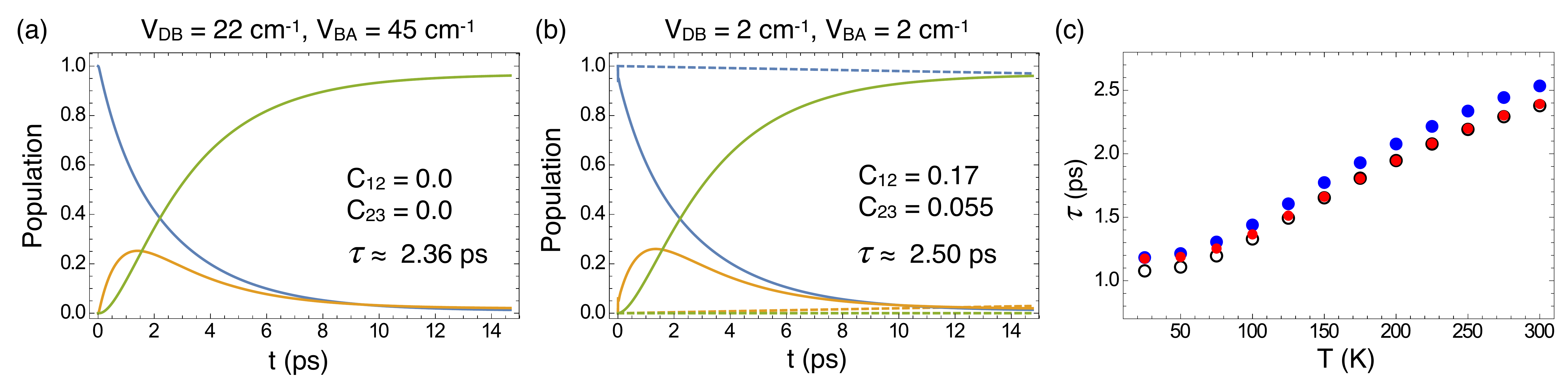}
\caption{
Dynamics of the sequential D-B-A model ($\epsilon_D=0.0$, $\epsilon_B=-150$ \cm and $\epsilon_A=-1000$ \cm) without (a) and with  off-diagonal system-environment interaction (b) at $T=300$ K.
Blue, orange and green lines represent donor $D$, bridge $B$ and acceptor  $A$ populations, respectively. Lifetime $\tau$ were obtained by an exponential fit of the donor $D$ population dynamics.
(c) Temperature dependence of  the donor $D$ population lifetime $\tau$. 
Black and red  symbols mark simulations without (a) and with off-diagonal system-environment interaction (b), respectively.  Blue symbols mark simulations  with off-diagonal system-environment interaction but increased  bridge acceptor couplings $V_{BA}=20$~\cm, see 
\BF{SM}, Fig.~S.1.
Simulation parameters are summarized in SI, Table~S.1-S.3\BF{, see also Fig.~S.3 for the impact of long-time system bath memory on the dynamics.}
}
\label{fig:SEQ}
\end{figure*}
\paragraph*{Sequential ET via a low-energy bridge ($\epsilon_D > \epsilon_B$).}
We start by briefly summarizing the well-known dynamics in the sequential  D-B-A model of multi-step electron transfer (Fig.~\ref{fig:SEQ}, $\epsilon_D=0$, $\epsilon_B=-150$~\cm and $\epsilon_A=-1000$~\cm) with diagonal system-bath interaction (cf. eq.~\ref{eq:SBM}, $d_1=2$, $d_2=1$, $d_3=0$ and $C_{12}=C_{23}=0$, simulation parameters are summarized in 
\BF{SM}, Table~S.1-S.3)
that resembles ET  in the bacterial RC.\cite{Sim:JPCB:1997,Fingerhut:CPL:2008,Fingerhut:JPCL:2012}
For the appreciable electronic couplings ($V_{DB}=22$~\cm $<$ $V_{BA}=45$~\cm) the charge transfer dynamics is characterized by low-picosecond donor $D$ to bridge  $B$ dynamics, followed by faster, sub-ps bridge $B$ to acceptor $A$ population transfer. The strong coupling to the environment ($\lambda_R$ = 501~\cm) is reflected in incoherent population dynamics. 
Despite the faster second transfer step, intermediate Bridge population $\approx$~25~\% is observed. The initial Donor-to-Bridge charge transfer dynamics  is characterized by an effective time constant $\tau \approx 2.36$~ps (single exponential fit). As expected, for reduced  electronic couplings  the charge transfer dynamics is substantially decelerated, now occurring on a $\approx$ 100 ps timescale (dashed lines in Fig.~\ref{fig:SEQ} (b), $V_{DB}=2$~\cm $\leq$ $V_{BA}=2-20$~\cm, see also 
\BF{SM}, Fig.~S.1).
%

Off-diagonal system-environment interaction can profoundly alter this scenario  (solid lines in  Fig.~\ref{fig:SEQ}~(b), $C_{12}$ $\neq$  0, $C_{23}$$\neq$ 0). We find that  the decay dynamics of the initially populated  donor state $D$ can get substantially accelerated once off-diagonal system-environment interactions are included. For the particular choice of $C_{12}$ and $C_{23}$, the well-known sequential dynamics of the D-B-A model is closely mirrored, 
even for small electronic coupling ($V_{DB}=2$ \cm and $V_{BA} = 2$ \cm, cf. eq.~\ref{eq:SBM}). The decay of the initially populated donor state $D$ is characterized by $\tau \approx 2.50$ ps.
%

Figure~\ref{fig:SEQ} (c) presents the temperature dependence of the donor decay lifetime $\tau$ in the range 25-300 K. 
For the investigated cases of diagonal and non-diagonal system-environment interaction we find a moderate enhancement of the transfer rate $k = 1/\tau$ upon  lowering of the temperature
with a ratio $k(300~\mathrm{K})/k(25~\mathrm{K}) \approx 0.49$.
The observed weak temperature dependence reflects pseudo-activationless electron transfer reactions\cite{Bixon:CPL:1989}
 being preserved for off-diagonal system environment interactions.  In this regime, nonadiabatic multiphonon electron transfer theory is applicable and the strong coupling to the medium vibrational motion dominates the transfer process.
The weak temperature dependence reflects the fact that the crossing of the nuclear potential surfaces along the reaction coordinate of the ET reaction occurs within the energy range  of the donor state vibrational levels
and medium reorganization is fast compared to the timescale of the transfer process. In the considered temperature range, the memory time $\tau_M < 250$ fs while the timescale of system dynamics is $\tau \sim [1.1 - 2.4]$ ps. 

\BF{The fast medium reorganization ($\tau_M \ll \tau$) indicates that the dynamics is  governed by a non-adiabatic, close-to-Markovian  process. We have investigated the influence of memory time $\tau_M$ on the transfer dynamics for constant reorganization energy $\lambda_R$ (Fig.~S.3). We find that for reduced memory time $\tau_M$ the donor-to-bridge transfer dynamics is moderately slowed down while the impact on the secondary bridge-to-acceptor transfer step is more pronounced. These findings resemble the findings from multi-state tight-binding models\cite{Richter:2017,Lambert:MolPhys:2012}: because intersite couplings are small compared to the energy gap, population dynamics is substantially affected  by the bath dynamics and the non-equilibrium bath state can  accelerate the population dynamics if long-time system bath memory time is taken into account ($\Delta k_{max} = 32$). Thus, the findings demonstrate the impact of non-Markovian long-time system-bath memory on charge transport dynamics in the considered overdamped transfer regime.}
%
%
%


An understanding of the acceleration of dynamics in the sequential D-B-A model despite the small electronic couplings $V_{ij}$ can be obtained by considering the system Hamiltonian $H'$ in transformed basis (eqs.~\ref{eq.9}-\ref{eq.10}). The unitary transformation $U$, determined  by off-diagonal system-bath interaction, induces substantial coupling elements $V_{DB}'$ and $V_{BA}'$ in intermediate, transformed basis, while approximately retaining the energetics ($\epsilon_i \approx \epsilon'_i$). 
As a consequence, \BF{we observe} acceleration of the sequential population dynamics induced by the off-diagonal system bath interaction.
The direct  $|D\rangle \to  |A\rangle$ transfer matrix element $V_{DA}'$ remains negligible compared to the donor-acceptor energy gap and, thus, the direct pathway  $|D\rangle \to  |A\rangle$ is irrelevant.
This mechanism is further corroborated by the moderate temperature dependence of the donor decay lifetime $\tau$ (Fig.~1(c)) that reflects the largely preserved energetics in transformed basis $\Delta E_{DB} \approx \Delta E_{DB}' \approx \lambda_R$. 
For the considered diagonal and off-diagonal system bath interaction cases, the largely preserved temperature dependence suggests that the dynamics proceeds via the sequential $|D\rangle \to |B\rangle \to |A\rangle$ pathway also for off-diagonal system-environment interaction.
%
%


%
%

\vspace{0.5cm}
\paragraph*{D-B-A Model in Asymmetric-$\Lambda$ Configuration ($\epsilon_D < \epsilon_B$,  $\epsilon_B - \epsilon_D \approx k_BT$).}
\label{sec:ALambda}
\begin{figure}[t!]
\includegraphics[width=.42\textwidth]{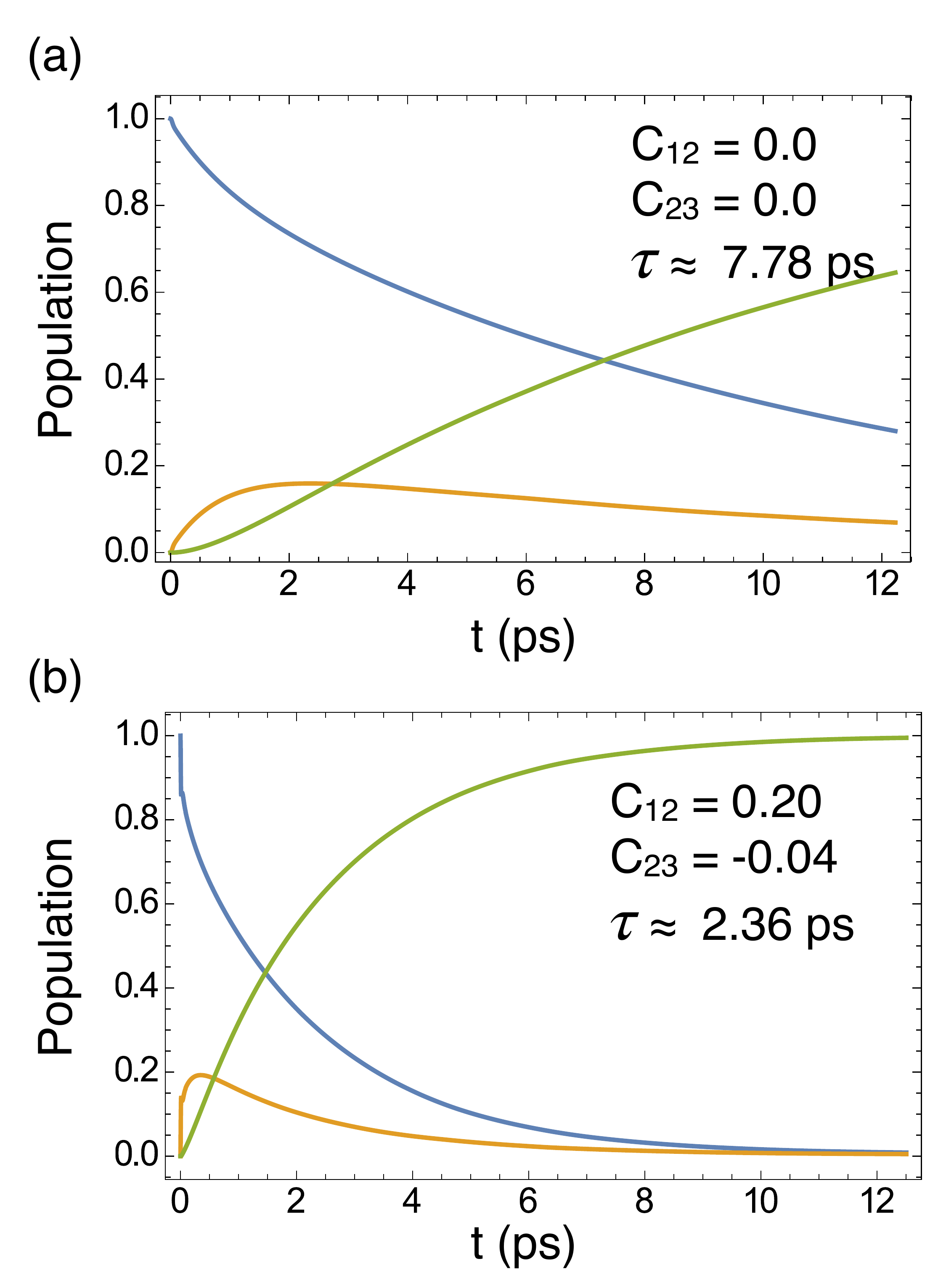}
\caption{
Dynamics of the  D-B-A model in asymmetric $\Lambda$  configuration ($\epsilon_D=0.0$, $\epsilon_B = + 200$~\cm $> \epsilon_D$ and $\epsilon_A=-1000$ \cm) without (a) and with (b )off-diagonal system-environment interaction ($V_{DB}=22$ \cm and $V_{BA}=45$ \cm).
Blue, orange and green lines represents the populations of donor state $D$, bridge state $B$ and acceptor state $A$, respectively.
The dynamics with diagonal and off-diagonal system-environment interaction for $\epsilon_B=300$ \cm are shown in the 
\BF{SM}, Fig.~S.2.
Simulation parameters are summarized in 
\BF{SM}, Table~S.1-S.3\BF{, see also Fig.~S.3 for the impact of long-time system bath memory on the dynamics.}
}
\label{fig:ALambda}
\end{figure}
\begin{figure*}[t!]
\includegraphics[width=1.\textwidth]{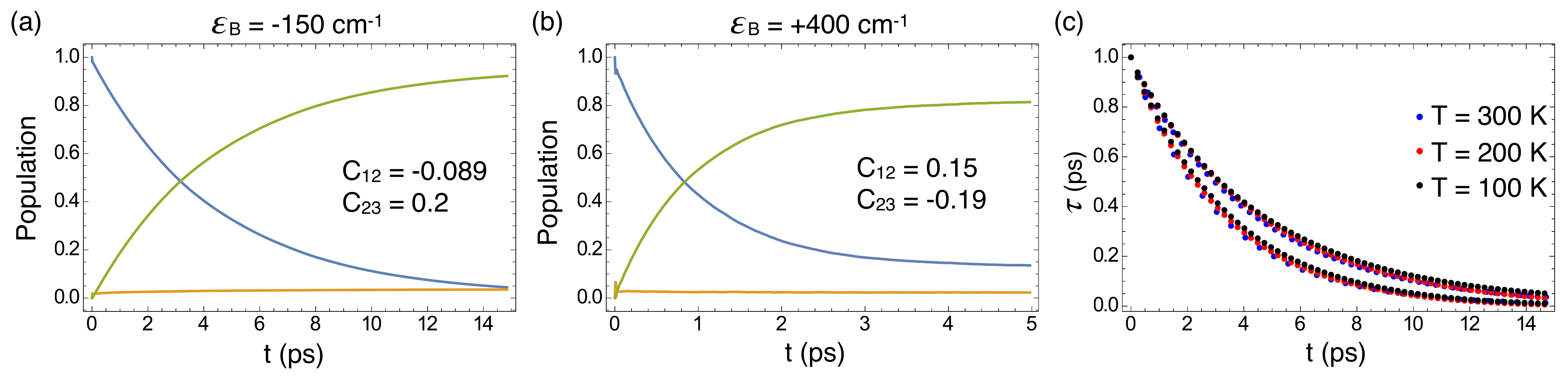}
\caption{
Off-diagonal induced superexchange transfer pathway $|D\rangle \to  |A\rangle$ for (a) the sequential D-B-A model  ($\epsilon_D=0.0$, $\epsilon_B=-150$ \cm and $\epsilon_A=-1000$ \cm)  and (b) the D-B-A model in asymmetric $\Lambda$ configuration ($\epsilon_D=0.0$, $\epsilon_B=+400$ \cm and $\epsilon_A=-500$ \cm).
Electronic couplings are $V_{DB}=22$ \cm and $V_{BA}=80$ \cm~(a) and  $V_{DB}=38$ \cm and $V_{BA}=90$ \cm~(b).
Blue, orange and green lines represents the populations of donor state $D$, bridge state $B$ and acceptor state $A$, respectively.
(c) Population decay of the donor  state $D$ for various $T$  in the setting of  the sequential D-B-A model with off-diagonal induced superexchange pathway (a,b).
%
Simulation parameters are summarized in 
\BF{SM}, Table~S.1-S.3.
%
%
%
}
\label{fig:SuEx}
\end{figure*}
Mutations allow to tune the energy of the bridge state in bacterial RC in a range  $-500$~\cm $< \epsilon_B < 1000$~\cm.\cite{Bixon:ChemPhys:1995}
The decay dynamics in the D-B-A Model with asymmetric-$\Lambda$ configuration with a thermally accessible bridge is presented in Fig.~\ref{fig:ALambda} (a) for diagonal interaction with the environment ($\epsilon_D < \epsilon_B = +200$~\cm; data for $\epsilon_B = +300$ are given in 
\BF{SM}, Fig.~S.2).
 The depopulation of the donor $D$
 is substantially decelerated compared to the sequential D-B-A model with $\epsilon_D > \epsilon_B$ due to thermal activation slowing down the transfer dynamics ($\tau \approx 7.78$ ps).
 For diagonal coupling to the environment, fast dynamics on a low-ps timescale (cf. Fig.~\ref{fig:SEQ}) can only be achieved by  substantially increased electronic couplings $V_{DB}$ and $V_{BD}$.\cite{Sim:JPCB:1997}


Considering  off-diagonal coupling to the environment ($C_{12} \neq 0, C_{23} \neq 0$,  Fig.~\ref{fig:ALambda} (b)), substantially accelerated dynamics can be realized compared to the diagonal interaction case and the dynamics can closely resemble the sequential ET dynamics via a low-energy bridge ($\epsilon_D > \epsilon_B$, cf. Fig~\ref{fig:SEQ}). In particular, the characteristic time constant for depopulation of the donor state is nearly indistinguishable ($\tau = 2.36$~ps).
 Despite the moderate $V_{DB}$ and $V_{BA}$, the off-diagonal component of the system-bath interaction can induce large electronic couplings $V_{DB}'$ and $V_{BA}'$ in the transformed basis while the impact of basis rotation on the energetics is minor.
 Similarly, the direct coupling element  $V_{DA}'$ in transformed basis is negligible for the chosen configuration $C_{12} = 0.2, C_{23} = -0.04$ ($V_{DA}' \lesssim 15$ \cm) 
 and the dynamics proceeds via the sequential pathway  $|D\rangle \to |B\rangle \to |A\rangle$. 
 Consequently, the dynamics in the asymmetric-$\Lambda$ D-B-A configuration with a thermally accessible bridge can potentially be fast in presence of  off-diagonal system-bath interactions.
 In this situation, the donor decay lifetime $\tau$ is insufficient to  distinguish   between  the energetics of the  asymmetric-$\Lambda$ model with off-diagonal interactions and sequential  ET proceeding via a low energy bridge state of the D-B-A system.

%
%
%

The parameters $d_1$, $d_2$ and $d_3$ represent the relative strength of the fluctuation of the energy levels owing to the bath interactions. On the other hand, $C_{12}$ and $C_{23}$ denote the strength of fluctuation of the electronic couplings $V_{DB}$ and $V_{BA}$, respectively. 
The simulations thus demonstrate that small off-diagonal contributions ($C_{ij} << d_1-d_2$ or $d_2-d_3$), i.e.,  non-Condon effects in $V_{DB}$,  can substantially increase the effective coupling element  and accelerate the dynamics. Here, the negative sign of $C_{12}$ and/or $C_{23}$ denotes the relative phase realized via anti-correlated fluctuations.
\BF{Notably, correlated site energy and transfer coupling fluctuations have been predicted for the FMO light harvesting complex while correlations of site energy fluctuations were found to be negligible.\cite{Olbrich:JPCB:2011}}
\subsection{Off-Diagonal Induced Superexchange Transfer Pathway $|D\rangle \to  |A\rangle$}
\label{sec:SuEx}
Variations of the D-B-A energetics in RC can be induced by point mutations that allow to alter the relative importance of the sequential $|D\rangle \to  |B\rangle \to  |A\rangle$ and the direct, superexchange mediated pathway  $|D\rangle \to  |A\rangle$. \cite{Bixon:ChemPhys:1995,Bixon:JCP:1997} For the latter, transient population of the bridge is avoided while the vibrational manifold of bridge state mediates the transfer.\footnote{We adopt the definition of Ref.~\citenum{Bixon:JCP:1997} where a superexchange mechanism is reflected in a vanishing  transient population of the bridge $B$. This definition differs from Ref.~\citenum{Skourtis:ChemPhys:1995} where  superexchange mechanism is mediated by the off-diagonal elements of the density matrix.}
%
In the following we consider  a sequential D-B-A model ($\epsilon_D=0$, $\epsilon_B=-150$~\cm, $\epsilon_A=-1000$~\cm) for which 
with diagonal system-bath interactions sequential  bridge mediated transfer is the only pathway due to absence of the direct coupling between donor $D$ and acceptor $A$ ($V_{DA}=0$, $V_{DB}=22$~\cm and 
\BF{$V_{BA}=80$}~\cm). 
In contrast, for  non-diagonal interaction with the environment ($C_{12}=-0.089
, C_{23}=0.2$) 
 the initially populated donor state $D$ decays directly  to the acceptor $A$. The negligible population of the bridge state $B$ (Fig.~\ref{fig:SuEx} (a)) suggests 
 that  
 the decay proceeds via a direct $|D \rangle \to |A\rangle$ channel, while the $|D\rangle \to |B\rangle \to |A\rangle$ channel is avoided. 
 A similar scenario is realized in the asymmetric-$\Lambda$ D-B-A  model with non-diagonal coupling ($\epsilon_B=400$\cm, Fig.~\ref{fig:SuEx}~(b)) where transfer via the $|D \rangle \to |A\rangle$ channel is dominant and population of the bridge state $B$ is avoided.
 
 The predominance of the direct  $|D \rangle \to |A\rangle$ channel is confirmed 
by the fact that the dynamics  is largely independent of the energy of the bridge state ( $\epsilon_B=-150$~\cm vs $\epsilon_B=400~$\cm, cf. Fig.~\ref{fig:SuEx} (a) and (b)). Further, the decay dynamics of the donor state $D$ is largely independent to variations in temperature ($T = $ 100 -300 K, Fig.~\ref{fig:SuEx} (c)), contrasting to the pseudo-activationless electron transfer reactions of the sequential D-B-A model (Fig.~\ref{fig:SEQ}).
In the latter scenario, the direct  $V_{DA}'$ coupling in transformed basis is negligible. The  dominant $|D \rangle \to |A\rangle$ channel in  presence of non-diagonal system-environment interaction can arise if in transformed basis if the $V_{DA}'$ coupling element becomes dominant while coupling element of the direct pathway  $V_{DB}'$ becomes suppressed. Accordingly,  the sequential population transfer via the bridge $B$ is negligible, while the dominant transfer channel occurs from the donor to the acceptor  $|D \rangle \to |A\rangle$.

\vspace{0.5cm}
\paragraph*{Competition between sequential $|D\rangle \to |B\rangle \to |A\rangle$  and superexchange $|D \rangle \to |A\rangle$ transfer pathways.}
\begin{figure}[t!]
\includegraphics[width=.42\textwidth]{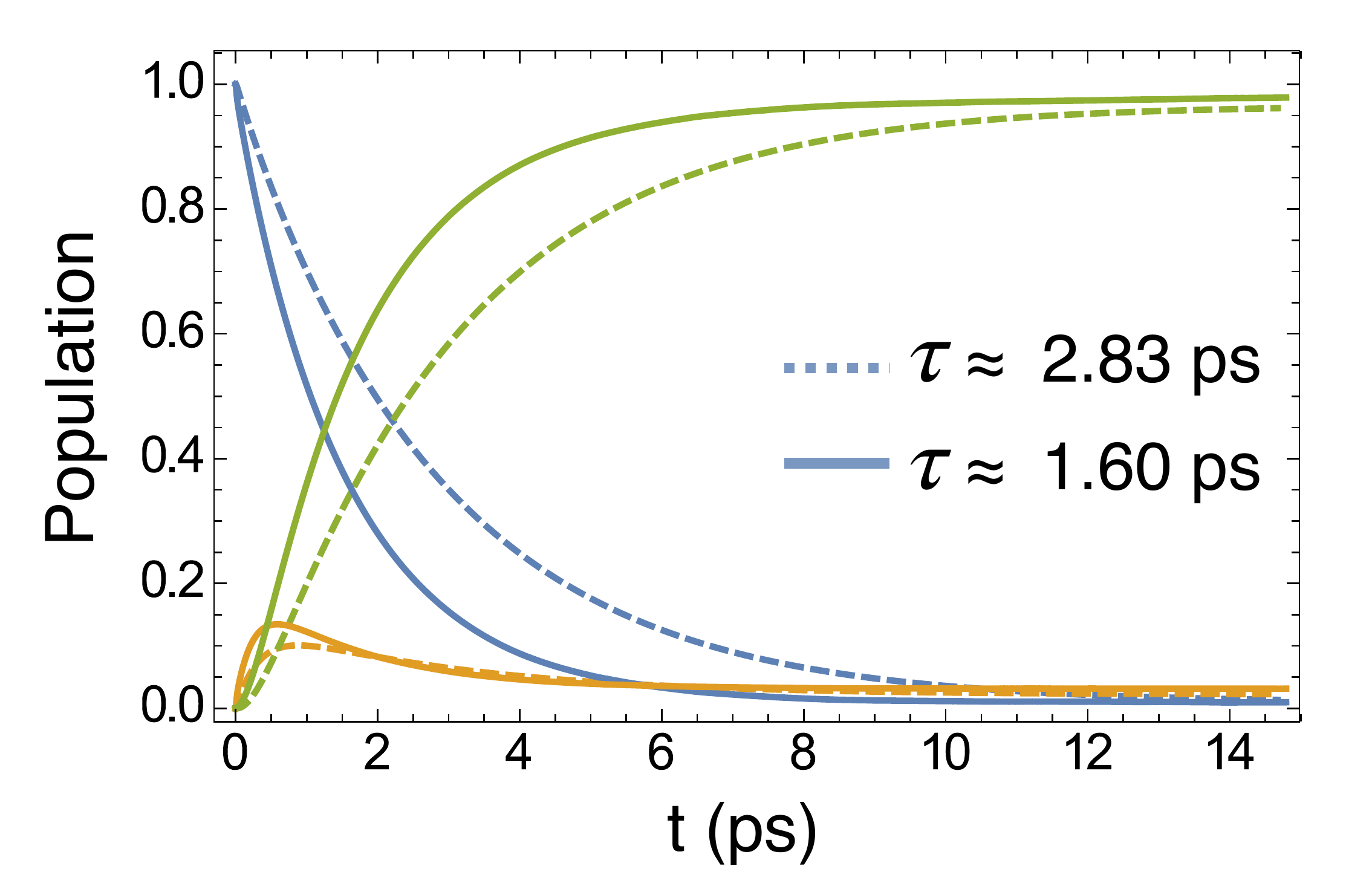}
\caption{
 Sequential $|D\rangle \to |B\rangle \to |A\rangle$  and super-exchange $|D \rangle \to |A\rangle$ pathway competition in the sequential  D-B-A model ($\epsilon_D=0.0$, $\epsilon_B=-150$ \cm and $\epsilon_A=-1000$ \cm). Dashed and solid lines denote simulations with diagonal   ($C_{12}= C_{23}=0$) and with off-diagonal  ($C_{12}=0.08, C_{23}=0.2$) system-environment interaction, respectively ($V_{DB}=22$~\cm and $V_{BA}=80$ \cm).
 Blue, orange and green lines represents the populations of donor state $D$, bridge state $B$ and acceptor state $A$, respectively.
 Simulation parameters are summarized in 
 \BF{SM}, Table~S.1-S.3\BF{, see also Fig.~S.3 for the impact of long-time system bath memory on the dynamics.}
 %
%
%
}
\label{fig:SeqSe}
\end{figure}
Starting from the generic sequential D-B-A model  ($\epsilon_D=0$, $\epsilon_B=-150$~\cm, $\epsilon_A=-1000$~\cm) we further investigate a scenario where both, the sequential  $|D\rangle \to |B\rangle \to |A\rangle$ and the superexchange mediated    $|D \rangle \to |A\rangle$ transfer pathway contribute simultaneously to the dynamics (Fig.~\ref{fig:SeqSe}). 
We find that the presence of off-diagonal system environment interactions
can activate both transfer pathways and the overall population decay of  the donor state $D$ becomes accelerated (characteristic time constant $\tau\approx 1.6$ ps 
compared to $\tau \approx 2.83$ ps for  the diagonal system-bath interaction case). 
The observed behavior can again be rationalized by considering the Hamiltonian in transformed basis (eqs.~\ref{eq.9}-\ref{eq.10}).
The particular unitary transformation induces couplings of comparable magnitude between all contributing states $D$, $B$ and $A$ ($V_{DB}' =-21.32 $~\cm, $V_{BA}' =-85.96 $~\cm, $V_{DA}' =114.24 $~\cm) while the  energetics of the sequential D-B-A model is largely preserved ($\epsilon_D'=-16.72$~\cm, $\epsilon_B= 147.1$~\cm and $\epsilon_A=-986.19$~\cm).
Due to the comparable magnitude of  coupling elements $V_{DA}'$ and  $V_{BA}'$,  the superexchange pathway  $|D \rangle \to |A\rangle$  now competes with the $|D\rangle \to |B\rangle \to |A\rangle$ pathway.
The off-diagonal system bath interaction here operates by mediating the additional direct $|D \rangle \to |A\rangle$ decay channel, while at the same time  the dynamics of the sequential $|D\rangle \to |B\rangle \to |A\rangle$ pathway is presered.

\subsection{Anomalous Bridge Localization}
\label{sec:Loc}
\begin{figure}[b!]
\includegraphics[width=.42\textwidth]{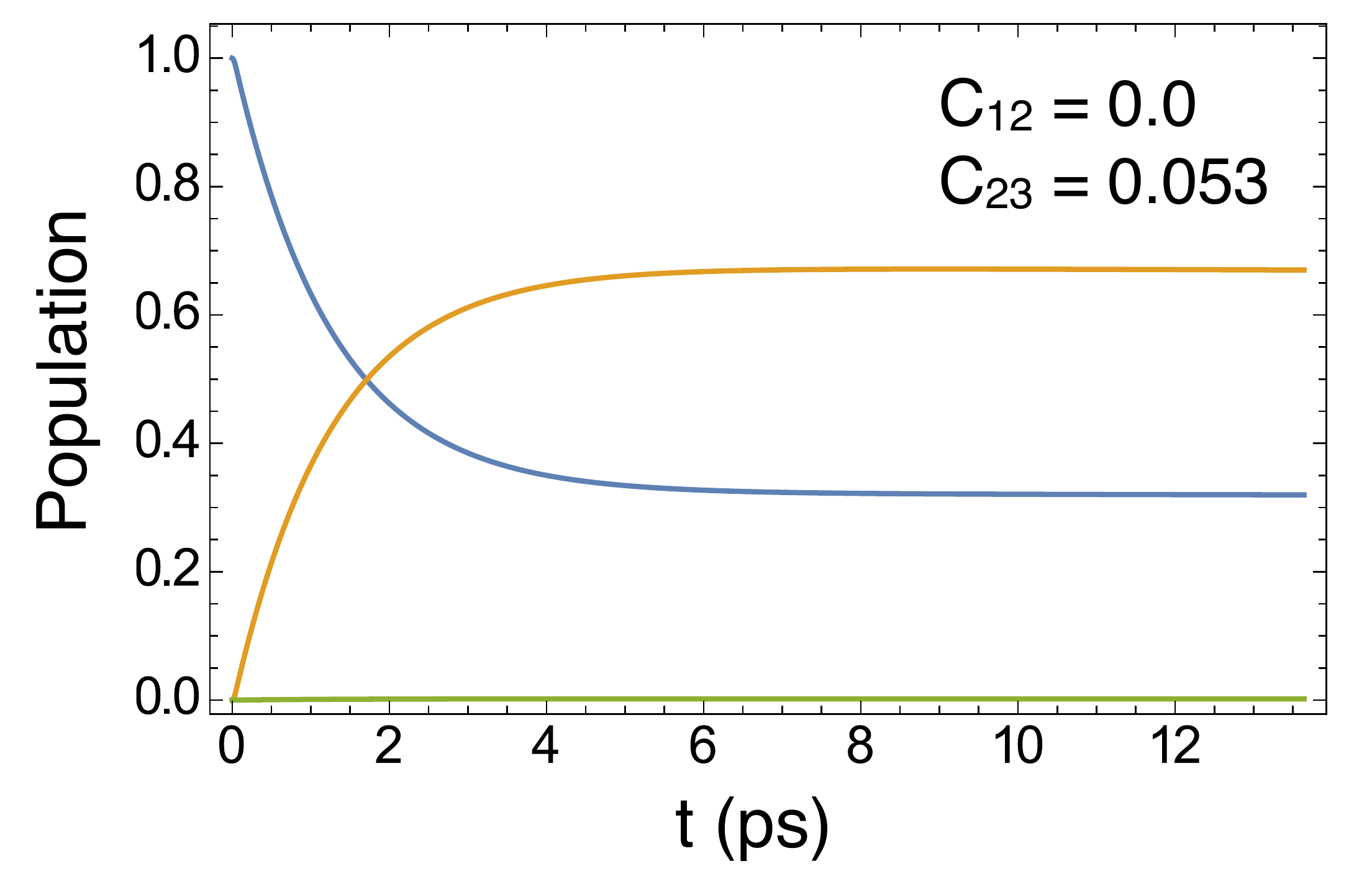}
\caption{
Anomalous localization in the sequential D-B-A model ($\epsilon_D=0.0$, $\epsilon_B=-150$ \cm and $\epsilon_A=-1000$ \cm) with electronic couplings $V_{DB}=22$ \cm and $V_{BA}=45$ \cm.
 Blue, orange and green lines represents the populations of donor state $D$, bridge state $B$ and acceptor state $A$, respectively.
  Simulation parameters are summarized in 
  \BF{SM}, Table~S.1-S.3.
%
%
}
\label{fig:loc1}
\end{figure}
%
%
%
%
In the following we discuss distinct scenarios of the sequential D-B-A model where off-diagonal system bath interactions induce anomalous localization of population in the bridge state $B$.

\vspace{0.5cm}
\paragraph*{Suppression of B-A Population Transfer.}
Anomalous localization in the bridge state $B$ can be induced if off-diagonal interaction primarily affects the second  $|B \rangle \to |A\rangle$ transfer step ($C_{12}=0.0 , C_{23}=0.053$, Fig.~\ref{fig:loc1}).
 %
Here, population transfer initially occurs from donor $D$ to bridge $B$ on the 2-3 ps timescale while the direct transfer pathway  $|D \rangle \to |A\rangle$ is negligible.
 However, subsequent $|B \rangle \to |A\rangle$  
  transfer is suppressed in the particular off-diagonal system environment configuration and population acquired in the bridge $B$ is thus unable to decay to the acceptor $A$.
 Figure~\ref{fig:loc1} demonstrates that such anomalous localization in the bridge $B$ can be realized with a small off-diagonal component particularly impacting the  $|B \rangle \to |A\rangle$ dynamics ($C_{23}\neq 0$).
 The anomalous localization  scenario is realized if  in the transformed basis  the coupling element $V_{BA}'$ is suppressed while $V_{DA}'$ is very small (compared to the  donor acceptor energy gap). 
 As the bridge-acceptor coupling $V_{BA}'$  is negligible, the population acquired in the bridge $B$ is unable to decay to the acceptor state and, in the steady state, the bridge remains populated, while no population transfer to the acceptor is observed. 
 

\vspace{0.5cm}
\paragraph*{Anomalous Bridge Localization via Renormalization of Tunneling Amplitudes.}
%
%
\begin{figure*}[t!]
\includegraphics[width=1.\textwidth]{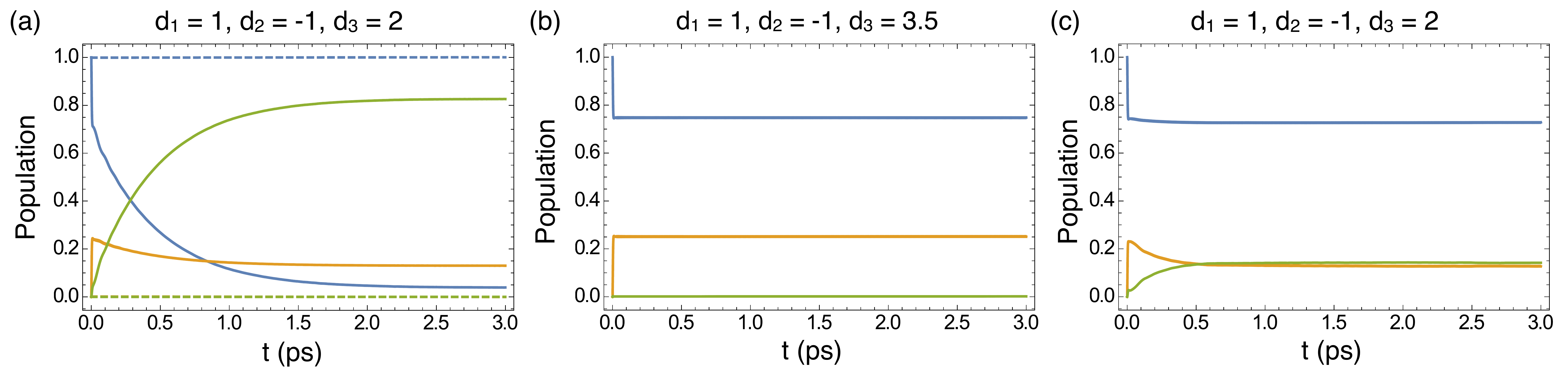}
\caption{
Off-diagonal induced anomalous bridge localization at low temperature ($T=0.05$ K) via renormalization of tunneling amplitudes $V_{ij} \rightarrow 0$ in the  sequential D-B-A model ($\epsilon_D=0$, $\epsilon_B=-100$ and $\epsilon_A=-500$, electronic couplings: $V_{DB} =V_{BA}=20$~\cm).
Dashed and solid lines in (a) denote the cases with diagonal and off-diagonal system-environment interactions, respectively. Solid lines show the  dynamics for the case $\alpha|D_1-D_2|>\alpha_c$ and $\alpha|D_2-D_3|>\alpha_c$ but $\alpha|D_1-D_3|<\alpha_c$ (with $D_i$ being the eigenvalues of the system-environment interaction matrix, eq.~\ref{eq:D}). 
(b) Dynamics for  the case $\alpha|D_1-D_2|>\alpha_c$, $\alpha|D_2-D_3|>\alpha_c$ and $\alpha|D_1-D_3|>\alpha_c$.
(c)  Dynamics for the 
\BF{case} 
$\alpha|D_1-D_2|>\alpha_c$ and $\alpha|D_1-D_3|>\alpha_c$ but $\alpha|D_2-D_3|<\alpha_c$.  
Blue, orange and green lines represents the populations of donor state $D$, bridge state $B$ and acceptor state $A$, respectively.
Simulation parameters are summarized in 
\BF{SM}, Table~S.1-S.3. 
%
}
\label{fig:BKT}
\end{figure*}
In the following we demonstrate population 
localization in the bridge state $B$  at low temperature where the underlying mechanism exploits the renormalization of the tunneling amplitudes for (ultra-)strong interaction with the environment. For diagonal system-environment interaction, the localization of a two-level system (Spin-Boson model) strongly interacting with the environment
is well understood.\cite{Weiss:book,Chakravarty:1982,Bray:1982,Strathearn:2018}
The renormalization of tunneling amplitudes $V_{ij} \rightarrow 0$ induces a freezing of the dynamics and thus a persistent localization of population 
that manifests in the Berezinskii-Kosterlitz-Thouless (BKT) 
 phase transition at the critical interaction strength $\alpha_c$. 
How such environment induced localization 
 is manifested  in presence of off-diagonal interactions with the environment for multi-level systems is largely unexplored. 
 
 Figure~\ref{fig:BKT} (a) shows the freezing of the population at the initial value due to tunneling amplitude
 renormalization induced by the strong diagonal interaction with the environment ($\alpha > \alpha_c \approx 1 +O(V_{DB}/\omega_c)$, dashed lines) at low temperature ($T=0.05$ K).
 The system is unable to decay via $|D \rangle \to |B \rangle \to |A \rangle$  if  $\alpha |d_1 -d_2| > \alpha_c$ is satisfied, as $V_{DB} \rightarrow 0 $. No condition for $d_3$ is required ($V_{DA}$ = 0, eq.~\ref{eq:Ham2}).
%
%
%

In presence of off-diagonal system-bath interactions, significant differences to the diagonal interaction case are identified.
The system exhibits  overdamped decay dynamics towards the thermal equilibrium state 
if $\alpha|D_1 -D_2| >\alpha_c$ and $\alpha|D_2 -D_3| >\alpha_c$ but $\alpha|D_1 -D_3| <\alpha_c$ (solid lines in Fig.~\ref{fig:BKT} (a), with $D_i$ being the eigenvalues of the system-environment interaction matrix, eq.~\ref{eq:D}).
Localization in the bridge $B$ via a freezing of dynamics due to a renormalization of tunneling amplitudes ($V_{ij} \rightarrow 0$) is only  observed 
if $\alpha|D_\alpha -D_\beta| >\alpha_c$ for all $\alpha \neq \beta$ ($\alpha,\beta=1,2,3$)~(Fig.~\ref{fig:BKT} (b)).
 In analogy to  the diagonal interaction case, the tunneling processes is suppressed due to tunneling amplitude renormalization but 
 initial short time dynamics persists that determines the degree of localization.\BF{\cite{Nirmal:2020}}
Figure~\ref{fig:BKT} (c) presents the dynamics for  $\alpha|D_1 -D_2| >\alpha_c$ and $\alpha|D_1 -D_3| >\alpha_c$ but $\alpha|D_2 -D_3| <\alpha_c$. 
In this scenario, population transfer between the donor $D$ and bridge $B$ is suppressed after the initial  short time dynamics but subsequent 
population transfer among bridge $B$ and the acceptor  state $A$ is facilitated.

In presence of off-diagonal system-bath interactions, we thus identify distinct mechanistic differences for the anomalous population localization  via the renormalization of tunneling amplitudes  $V_{ij} \rightarrow 0$:
%
(i) due to the basis transformation of initial conditions (eq.~\ref{eq:init}), coherences are 
imposed in the  initial conditions that induce short time dynamics and consequently a population of all 
 basis states $|D' \rangle$, $|B'\rangle$ and $|A' \rangle$;
(ii) stricter requirements are 
to be satisfied by the system-bath interactions in order to realize anomalous bridge localization due to a freezing of the dynamics, arising from the strong interaction in the BKT localized phase. For example, the condition for the diagonal interaction case, $\alpha|D_1-D_2|> \alpha_c$ with arbitrary $D_3$, is not sufficient because this only renormalizes $V_{DB}'$ to zero but not $V'_{DA}$  whose magnitude can become significant due to basis rotation. 
Thus, in presence of non-diagonal system-bath interactions, the conditions for bridge localization are stricter:
$\alpha|D_1 -D_2| >\alpha_c$, $\alpha|D_2 -D_3| >\alpha_c$ and $\alpha|D_1 -D_3| >\alpha_c$ need to be simultaneously satisfied to sufficiently renormalize $V_{BA}'$ and $V_{DA}'$.
Alternatively, when $\alpha|D_1 -D_2| >\alpha_c$ and $\alpha|D_1 -D_3| >\alpha_c$ but $\alpha|D_2 -D_3| <\alpha_c$ (Fig.~\ref{fig:BKT} (c)), the donor population freezes subsequent to  the initial short-time dynamics as 
both $V_{DB}'$ and $V_{DA}'$ are renormalized to zero, which suppresses tunneling processes involving the donor state $D$ but still facilitates equilibration between the bridge $B$ and acceptor $A$. 

\subsection{Donor-Acceptor Coherence Transfer}
\label{sec:Coh}
%
%
\begin{figure}
\includegraphics[width=0.39\textwidth]{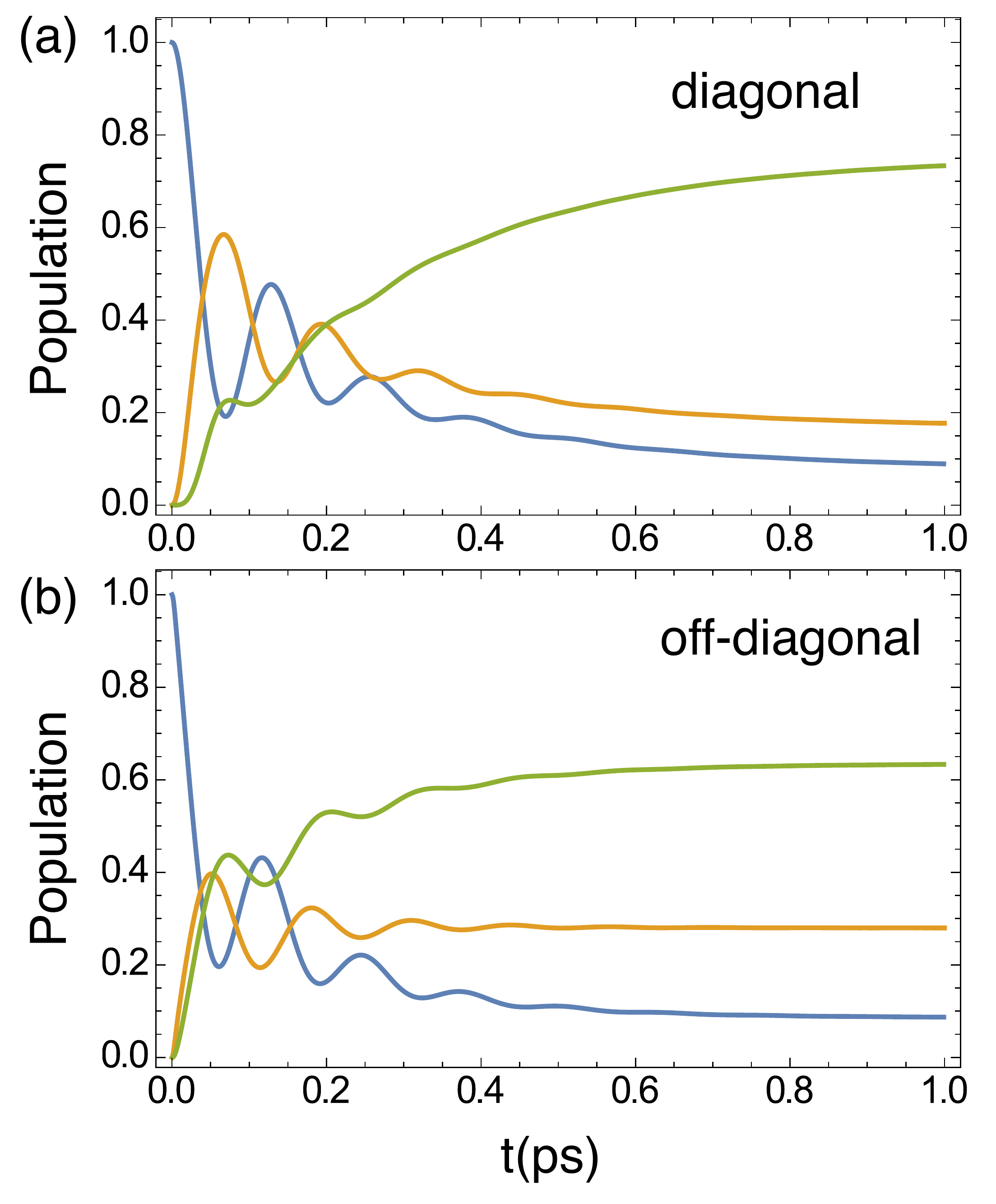}
    \caption{Donor-acceptor coherence transfer  in a sequential D-B-A model ($\epsilon_D = 0$ \cm, $\epsilon_B = -100$ \cm, $\epsilon_A = -500$ \cm; electronic couplings: $V_{\rm DB}=130$ \cm, $V_{\rm BA}=145$ \cm ) at room temperature ($T=300$ K). The real time dynamics is shown for diagonal (a) and off-diagonal (b) system-environment interactions (system-environment interaction strength $\alpha$ = 0.835).
     Off-diagonal system-environment interactions are $C_{12} = 0.2$, $C_{23} = 0.9$ (b).
    Blue, orange and green lines represents the populations of donor state $D$, bridge state $B$ and acceptor state $A$, respectively.
Simulation parameters are summarized in 
\BF{SM}, Table~S.1-S.3. 
%
}
    \label{fig:coh1}
\end{figure}

%
%
The prototypical dynamics in the sequential D-B-A model (Fig.~\ref{fig:SEQ}) \BF{proceeds} overdamped and 
\BF{is} reasonably accounted for by pseudo-activationless, non-adiabatic ET theory.\cite{Bixon:CPL:1989,Huppmann:JPCA:2003,Fingerhut:CPL:2008}
Accordingly, the timescales of population transfer and environment relaxation are largely separated and non-Markovian environment relaxation has a modest impact on donor $D$ and transient bridge  $B$ population \BF{(cf. SI, Fig.~S.3)}.\cite{Sim:JPCB:1997}
 \BF{Nevertheless, in biological ET, where the sluggish  protein matrix together with water fluctuations are of paramount importance, a close interplay between energy  gap and transfer coupling element fluctuations is anticipated.\cite{Skourtis:AnnRevPhysChem:2010} Such interplay  is particularly permissible if the correlation time of site energies and transfer couplings is comparable to the time scale of the ET process.}
\BF{In our simulations of the transfer dynamics in the  prototypical D-B-A
model (Sec.~\ref{sec:seq} and Sec.~\ref{sec:SuEx}), energetics and magnitude of transfer couplings are chosen to resemble the dynamics of the bacterial RC\cite{Sim:JPCB:1997,Fingerhut:CPL:2008,Fingerhut:JPCL:2012} for which details of site energy and  transfer coupling correlations have not been reported.}
\BF{Correlations of site energy and  transfer coupling fluctuations have been proposed for the FMO light harvesting complex \cite{Olbrich:JPCB:2011} and non-Condon fluctuations of transfer couplings have been reported, e.g., for the PE545 light harvesting complex\cite{Aghtar:JPCB:2017} and for ET at oligothiophene-fullerene interfaces\cite{Tamura:JCP:2012} where reported magnitudes of off-diagonal fluctuation are  consistent with correlated off-diagonal system-environment interactions $|C_{ij}|\lessapprox$ 0.2.} 
 %

 \BF{In fact, t}he short-time pump-probe and 2D spectra of bacterial RC\cite{Vos:Nature:1993,Sporlein:JPCB:1998,Novoderezhkin:JPCB:2004,Ma:NatCom:2019}
 show coherent, non-exponential dynamics that potentially arises from (ground or excited state) vibrational wavepackets, indicating comparable timescales of population transfer and environment relaxation. 
 Of particular interest 
 are 
 coherent signatures at the spectral position indicative of the acceptor $A$ (bacteriopheophytin $H_A$) \BF{that indicate the possibility of direct donor-to-acceptor coherence transfer}.
%
  %
  Recent 2D measurements 
  \BF{suggest} a \BF{scenario with} more complex electronic and vibrational interactions 
  within the first hundreds of femtosecond\BF{s} in bacterial RC\cite{Ma:NatCom:2019}.

 \begin{figure}[b!]
 \includegraphics[width=0.39\textwidth]{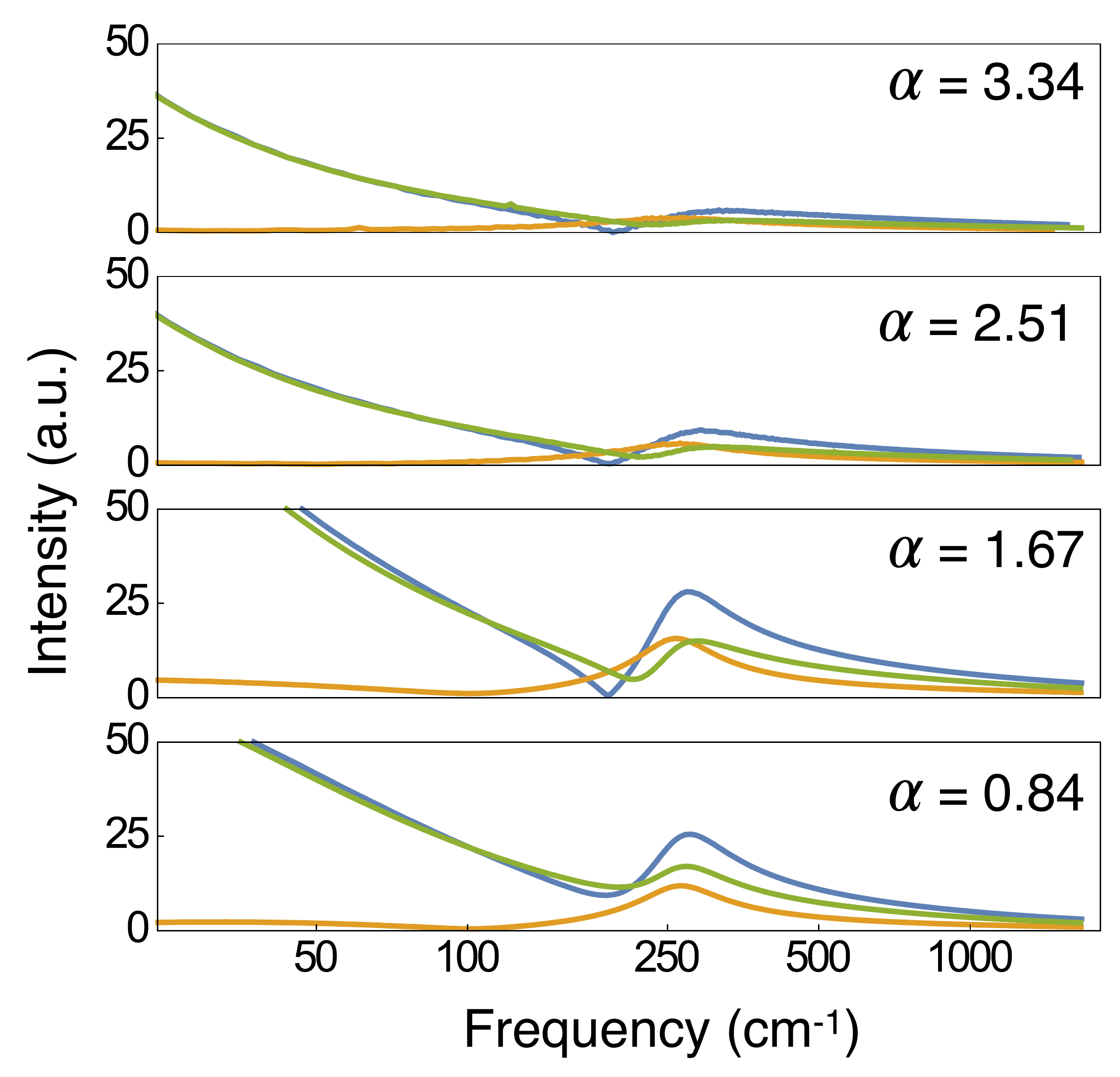}
    \caption{
    Frequency domain representation of the oscillatory dynamics in a sequential D-B-A model for various values of the system-environment interaction strength $\alpha$. Frequency domain data were obtained upon Fourier transform of the oscillatory dynamics for off-diagonal system-environment interactions ( $C_{12} = 0.2$, $C_{23} = 0.9$, see Fig.~\ref{fig:coh1}) after subtraction of the long-time equilibrium populations  of respective states.
  Blue, orange and green lines represents the populations of donor state $D$, bridge state $B$ and acceptor state $A$, respectively.  
    }
    \label{fig:coh2}
\end{figure}
 In the following,  we explore possibilities of  coherent transfer dynamics between donor $D$ and acceptor $A$ 
 and the impact of non-diagonal system-environment interactions.  
 Figure~\ref{fig:coh1} presents the dynamics of a modified sequential D-B-A model with
  increased donor-bridge and bridge\BF{-}acceptor coupling elements  ($V_{\rm DB}=130$ \cm, $V_{\rm BA}=145$ \cm). For ease of numerical efficiency we consider a reduced donor-acceptor energy gap ($\Delta (\epsilon_D - \epsilon_A) = 600$~\cm, $\epsilon_B$=-100~cm$^{-1}$).
  In the modified setting, accelerated transfer dynamics occurs
  and the timescales of population transfer and environment relaxation become comparable.
   Accordingly,
  %
 the population dynamics of the sequential D-B-A model shows coherent oscillatory modulations
  involving donor $D$, bridge $B$ and acceptor $A$ with 
   diagonal system-environment interaction (Fig.~\ref{fig:coh1} (a), $C_{12} = C_{23} = 0$). Oscillatory dynamics occurs on the few-hundred fs timescale and is followed by 
   exponential equilibration dynamics occurring on the $\approx$ 1 ps timescale. Oscillations appear particularly pronounced in the donor and bridge population while the acceptor population follows a step-function increase during the first $\approx$ 300 fs.  

Off-diagonal system-environment interactions allow to modulate the relative amplitudes  of the oscillatory donor, bridge and acceptor population dynamics (Fig.~\ref{fig:coh1} (b), $C_{12} = 0.2$, $C_{23} = 0.9$). In the particular realization, the  oscillatory decay of the donor state $D$ is largely preserved while the oscillation amplitude of the bridge state $B$ is decreased and the oscillation amplitude of the acceptor state $A$ is increased.
%
Figure~\ref{fig:coh2} presents  the frequency domain representation of the dynamics 
for varying  system-environment interaction strength $\alpha$.
We find that for increasing $\alpha$
the non-zero frequency components (centered around $\approx 300$~\cm) become weaker while   the zero frequency component increases, as expected for the transition from oscillatory to overdamped dynamics.
While the  peak position of the bridge state $B$ is largely unaffected, the peak positions characterizing the coherent dynamics of the donor $D$ and acceptor $A$ states  are slightly shifted to higher frequencies for increasing system-environment interaction strength $\alpha$.
%
The observed oscillation frequency is attributed (within $\approx$ 25~\cm) to the donor $D$-bridge $B$ energy gap of the diagonalized Hamiltonian in transformed basis (253~\cm). For weak-to-moderate interaction with the environment ($\alpha = 0.8-1.6$) the oscillation frequency appears nearly unperturbed by the interaction with bath  while for strong system-bath interaction deviations from the system Hamiltonian eigenvalues become larger (up to 60~\cm)  due to a bath induced renormalization of system frequencies. Thus, upon increase of the system-environment interaction strength $\alpha$ (Fig.~\ref{fig:coh2}) the coherence frequency monotonically shifts  away from the system resonance frequency to higher frequencies and the amplitude of coherences is reduced due to increased interaction strength with the environement.

\vspace{0.5cm}
\paragraph*{Donor-Acceptor Coherence Transfer in Absence of  Electron Transfer Couplings.}
%
%
\begin{figure}[t!]
\includegraphics[width=.42\textwidth]{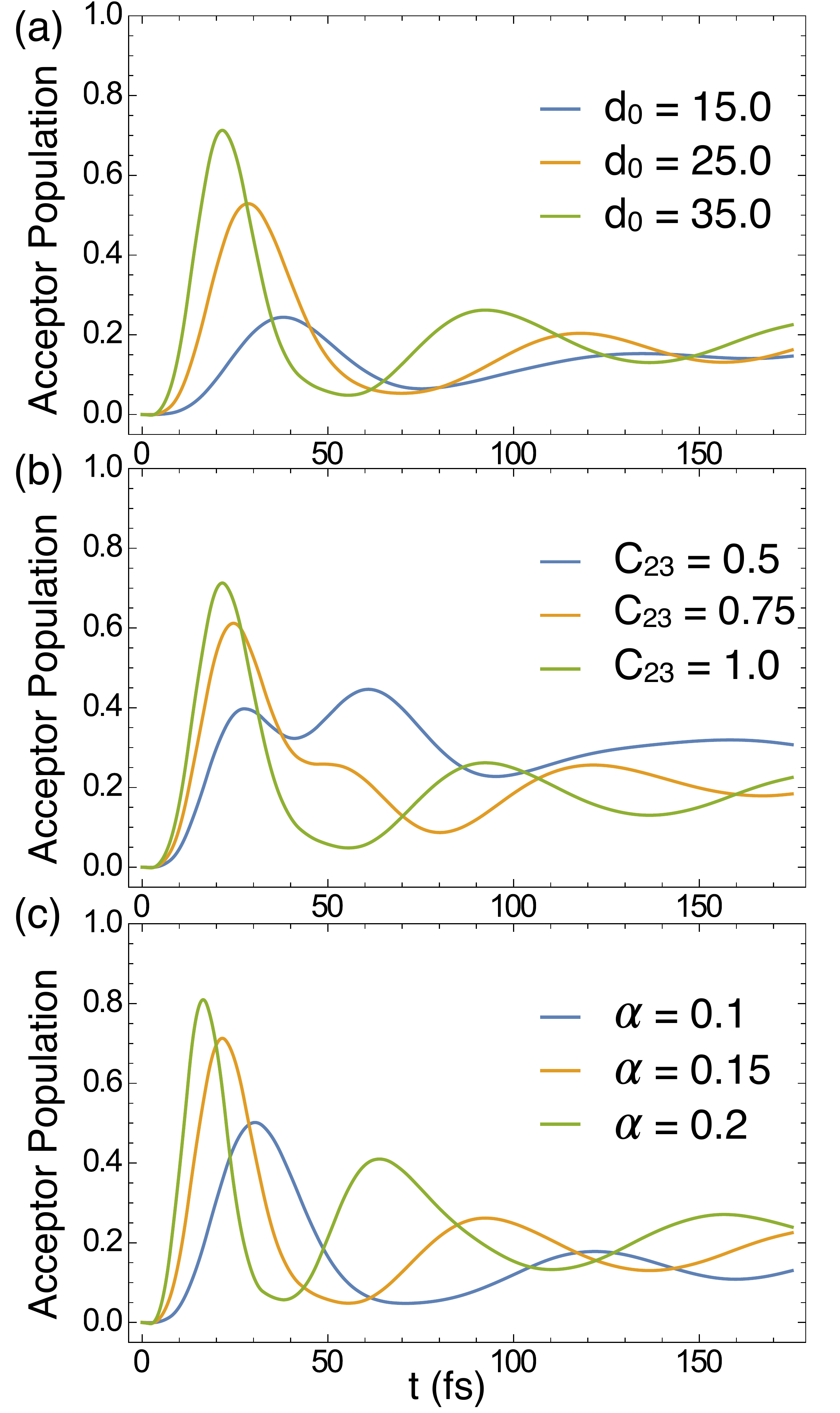}
\caption{
Donor-acceptor coherence transfer in the sequential D-B-A model ($\epsilon_D=0$, $\epsilon_B=-100$ \cm and $\epsilon_A=-500$ \cm, electronic coupling: $V_{DB}=V_{BA}$ = 0) at  strong system-environment interaction.
(a) Oscillatory dynamics for variations of the bath displacement  parameter $d_0$ ($\alpha = 0.15$, $C_{12} = C_{23} = 1$).
(b) Oscillatory dynamics for variation of the off-diagonal system-environment interaction  $C_{23}$ ($\alpha = 0.15$, $C_{12} = 1$,  $d_0 = 35$).
(c) Oscillatory dynamics for variation of the system-environment interaction strength $\alpha$  ($C_{12} = C_{23} = 1$,  $d_0 = 35$).
Diagonal entries of 
the system-environment interaction matrix are identical to the shift $d_0$ ($d_1=d_2=d_3=d_0$), simulation parameters are summarized in 
\BF{SM}, Table~S.1-S.3.
%
%
%
%
}
\label{fig:BKTdyn}
\end{figure}

%
%
%
%
In the following we numerically demonstrate that donor $D$ to acceptor $A$ coherence transfer can be induced via the off-diagonal interaction with the environment.
For purpose of demonstration we consider a sequential D-B-A model ($\epsilon_D=0$,  $\epsilon_B=-100$ \cm and $\epsilon_A=-500$ \cm) that neglects the electronic coupling elements which mediate the (coherent) transfer dynamics ($V_{DB} = V_{BA} = 0$). Thus, for diagonal system-bath interaction, the model does not show any dynamical time evolution.

Figure~\ref{fig:BKTdyn} demonstrates that in presence of   off-diagonal system-bath coupling ($C_{12}\neq  C_{23} \neq 0$, $d_0 \neq 0$,  $d_1=d_2=d_3 = d_0$) the short-time dynamics can become oscillatory.
 Most importantly, the simulations reveal  that not only the donor $D$ and bridge $B$ populations, but also the acceptor population shows  coherent dynamics. 
 The characteristic features of the oscillatory dynamics are distinct from coherent dynamics appearing at weak-to-intermediate coupling to the environment (cf. Fig.~\ref{fig:coh1}).
 We find that the oscillation frequency and the amplitude of oscillations increase with growing values of the coupling strength $\alpha$, the oscillator displacement $d_0$ and 
  the magnitude of off-diagonal system-environment interactions $C_{ij}$ (Fig.~\ref{fig:BKTdyn} (a-c)).  
  
%

The oscillatory  dynamics depicted in Fig.~\ref{fig:BKTdyn} (a) reveals that the diagonal displacement parameter $d_0$ plays a crucial role in determining the short-time dynamics. This finding contrasts the dynamics in the D-B-A model with  diagonal system-environment interaction  ($C_{12}=C_{23} = 0$) where a constant diagonal shift $d_0$ of the coupling matrix $d_1 \to d_1 +d_0$,  $d_2 \to d_2 +d_0$ and $d_3 \to d_3 +d_0$ does not affect the dynamics because the diagonal shift neither affects the coupling strength nor the tunneling amplitudes.\cite{Weiss:book}
In the DBA model with non-diagonal system-bath interactions, the non-exponential short-time dynamics proceeding equilibration via a tunneling processes is accelerated by the diagonal shift $d_0$ and oscillatory dynamics can be more pronounced (cf. also initial dynamics in Fig.~\ref{fig:ALambda}). 

%
%

The anomalous characteristics of the sort-time dynamics can be understood from the canonical mapping between the  D-B-A model with off-diagonal system-environment interactions and the vibronic interactions in the primary  reaction coordinate model (Sec.~\ref{sec:prc}, eq.~\ref{eq:HamPRC}). The non-equilibrium electronic state couples (non-diagonally) to the primary  mode that subsequently dissipates into a bath \cite{Chernyak:1996}.
The finite non-Markovian relaxation timescale of the environment here creates a bottleneck for energy dissipation giving rise to the observed short-time coherences.
  The non-equilibrium excitation of the primary mode can be transferred back to electronic state that can become susceptible to re-excitation
 which  leads to the observed oscillatory coherent dynamics in the D-B-A model in absence of electronic coupling elements. 
Accordingly,  the ultrafast short-time dynamics is dominated by quantum effects and has negligible tunneling contributions. 
The observed short-time oscillation frequency  becomes transparent
in the primary reaction coordinate framework: the timescale of the short-time dynamics and hence, the oscillations is determined by the energy scale $\omega_c$. In the primary reaction coordinate model, $\omega_c$ is connected to the frequency of the primary oscillator $\Omega$ and its damping coefficient $\gamma$ due to subsequent dissipation into the bath: $\omega_c =\Omega^2/\gamma$. 
For example, in the considered overdamped regime, a primary mode $\Omega \sim 500 $~\cm~with $\gamma \sim 2500$ \cm~yields $\omega_c=100$ \cm. 
A large oscillator displacement $d_0$ and off-diagonal coupling strength $C_{ij}$ accelerates the transfer of excitation from the electronic state to the primary mode thus enhancing non-equilibrium oscillatory vibronic features. 
The observed coherences (Fig.~\ref{fig:BKTdyn}) have frequencies on the order  $\sim 400-800$ \cm~that persist for few hundreds of femtoseconds. 
The off-diagonal induced electronic coherent process thus survives well on the time scale of molecular vibrations and is readily observable with available spectroscopic detection methods.\cite{Liebel:JPCA:2015,Kowalewski:ChemRev:2017}

\section{Conclusions}
We have numerically demonstrated in non-perturbative QUAPI simulations that non-diagonal system-environment interactions can profoundly affect the observed dynamics in the prototypic D-B-A model.
Off-diagonal induced acceleration of the dynamics has been demonstrated for the sequential D-B-A model with a low-energy bridge state and in asymmetric-$\Lambda$ configuration. Additionally, we have shown how off-diagonal system-environment interactions can activate the direct superexchange transfer pathway.
In all cases the subtle deviations from cases with diagonal system-environment interaction preclude the assignment of a definitive transfer mechanism from the dynamics alone.
 
 The suppression of bridge $B$ to acceptor $A$ population transfer mediated by the off-diagonal interaction with the environment allows for  anomalous localization of population in bridge states. 
 Two mechanisms of anomalous bridge $B$ localization are  identified where (i) due to basis transformation, the effective coupling element are suppressed and (ii)  anomalous localization in the bridge state can be induced at low temperature
via a renormalization of tunneling amplitudes for strong interaction with the environment. 
In the latter case, more stringent criteria for localization are identified compared to the diagonal case via the underlying localization mechanism  of the Spin-Boson model at strong interaction with the environment.

Coherence transfer between donor $D$ and acceptor $A$ was demonstrated to be affected by the  presence of off-diagonal system-environment interactions.
The environment mediated  $D$ - $A$ coherence transfer in absence of electron transfer couplings exhibits anomalous characteristics for the sort-time dynamics, i.e., an increase of the oscillation amplitudes with increasing system-environment interaction  strength, oscillator displacement and off-diagonal system-environment interactions. The characteristics become transparent upon the canonical mapping of the D-B-A model Hamiltonian with off-diagonal system-environment interactions onto the vibronic primary reaction coordinate model. 
 Our results reveal a novel view on   short-time coherent dynamics that arises from the complex and strong interaction of the environment  and electronic degrees of freedom induced by vibronic non-Condon effects.

\BF{\section*{Supplementary Material}}

\BF{Auxiliary dynamics of the sequential D-B-A model (Fig.~S.1) and  auxiliary dynamics of the  D-B-A model in asymmetric $\Lambda$  configuration (Fig.~S.2); dependence of dynamics 
on memory times $\tau_M$ (Fig.~S.3);
Tables S.1-S.3 giving parameters of system Hamiltonians, bath and interaction  Hamiltonians and MACGIC-QUAPI settings used in simulations as well as numerical demonstration of convergence (Fig. S.4).}
\vspace{.5cm}

\begin{acknowledgments}
B. P. F. acknowledges support by the DFG within the Emmy-Noether Program (Grant No. FI 2034/1-1). 
This research has received funding from the European Research Council (ERC) under the European Union’s Horizon 2020 research and innovation program (grant agreement No. 802817).
\end{acknowledgments}

\section*{Data Availability}
The data that support the findings of this study are available from the corresponding author upon reasonable request.

\bibliography{DBA_OffDiagonal_Quapi}

\begin{thebibliography}{58}%
\makeatletter
\providecommand \@ifxundefined [1]{%
 \@ifx{#1\undefined}
}%
\providecommand \@ifnum [1]{%
 \ifnum #1\expandafter \@firstoftwo
 \else \expandafter \@secondoftwo
 \fi
}%
\providecommand \@ifx [1]{%
 \ifx #1\expandafter \@firstoftwo
 \else \expandafter \@secondoftwo
 \fi
}%
\providecommand \natexlab [1]{#1}%
\providecommand \enquote  [1]{``#1''}%
\providecommand \bibnamefont  [1]{#1}%
\providecommand \bibfnamefont [1]{#1}%
\providecommand \citenamefont [1]{#1}%
\providecommand \href@noop [0]{\@secondoftwo}%
\providecommand \href [0]{\begingroup \@sanitize@url \@href}%
\providecommand \@href[1]{\@@startlink{#1}\@@href}%
\providecommand \@@href[1]{\endgroup#1\@@endlink}%
\providecommand \@sanitize@url [0]{\catcode `\\12\catcode `\$12\catcode
  `\&12\catcode `\#12\catcode `\^12\catcode `\_12\catcode `\%12\relax}%
\providecommand \@@startlink[1]{}%
\providecommand \@@endlink[0]{}%
\providecommand \url  [0]{\begingroup\@sanitize@url \@url }%
\providecommand \@url [1]{\endgroup\@href {#1}{\urlprefix }}%
\providecommand \urlprefix  [0]{URL }%
\providecommand \Eprint [0]{\href }%
\providecommand \doibase [0]{http://dx.doi.org/}%
\providecommand \selectlanguage [0]{\@gobble}%
\providecommand \bibinfo  [0]{\@secondoftwo}%
\providecommand \bibfield  [0]{\@secondoftwo}%
\providecommand \translation [1]{[#1]}%
\providecommand \BibitemOpen [0]{}%
\providecommand \bibitemStop [0]{}%
\providecommand \bibitemNoStop [0]{.\EOS\space}%
\providecommand \EOS [0]{\spacefactor3000\relax}%
\providecommand \BibitemShut  [1]{\csname bibitem#1\endcsname}%
\let\auto@bib@innerbib\@empty
\bibitem [{\citenamefont {Warshel}\ and\ \citenamefont
  {Schlosser}(1981)}]{Warshel:PNAS:1981}%
  \BibitemOpen
  \bibfield  {author} {\bibinfo {author} {\bibfnamefont {A.}~\bibnamefont
  {Warshel}}\ and\ \bibinfo {author} {\bibfnamefont {D.~W.}\ \bibnamefont
  {Schlosser}},\ }\href {\doibase 10.1073/pnas.78.9.5564} {\bibfield  {journal}
  {\bibinfo  {journal} {Proc. Nat. Acad. Sci. USA}\ }\textbf {\bibinfo {volume}
  {78}},\ \bibinfo {pages} {5564} (\bibinfo {year} {1981})}\BibitemShut
  {NoStop}%
\bibitem [{\citenamefont {Fingerhut}, \citenamefont {Zinth},\ and\
  \citenamefont {de~Vivie-Riedle}(2010)}]{Fingerhut:PCCP:2010}%
  \BibitemOpen
  \bibfield  {author} {\bibinfo {author} {\bibfnamefont {B.~P.}\ \bibnamefont
  {Fingerhut}}, \bibinfo {author} {\bibfnamefont {W.}~\bibnamefont {Zinth}}, \
  and\ \bibinfo {author} {\bibfnamefont {R.}~\bibnamefont {de~Vivie-Riedle}},\
  }\href {\doibase 10.1039/B914552D} {\bibfield  {journal} {\bibinfo  {journal}
  {Phys. Chem. Chem. Phys.}\ }\textbf {\bibinfo {volume} {12}},\ \bibinfo
  {pages} {422} (\bibinfo {year} {2010})}\BibitemShut {NoStop}%
\bibitem [{\citenamefont {Bixon}\ and\ \citenamefont
  {Jortner}(1997)}]{Bixon:JCP:1997}%
  \BibitemOpen
  \bibfield  {author} {\bibinfo {author} {\bibfnamefont {M.}~\bibnamefont
  {Bixon}}\ and\ \bibinfo {author} {\bibfnamefont {J.}~\bibnamefont
  {Jortner}},\ }\href {\doibase 10.1063/1.474878} {\bibfield  {journal}
  {\bibinfo  {journal} {J. Chem. Phys.}\ }\textbf {\bibinfo {volume} {107}},\
  \bibinfo {pages} {5154} (\bibinfo {year} {1997})}\BibitemShut {NoStop}%
\bibitem [{\citenamefont {Nitzan}(2001)}]{Nitzan:AnnRevPhysChem:2001}%
  \BibitemOpen
  \bibfield  {author} {\bibinfo {author} {\bibfnamefont {A.}~\bibnamefont
  {Nitzan}},\ }\href {\doibase 10.1146/annurev.physchem.52.1.681} {\bibfield
  {journal} {\bibinfo  {journal} {Ann. Rev. Phys. Chem.}\ }\textbf {\bibinfo
  {volume} {52}},\ \bibinfo {pages} {681} (\bibinfo {year} {2001})}\BibitemShut
  {NoStop}%
\bibitem [{\citenamefont {Goldsmith}, \citenamefont {Wasielewski},\ and\
  \citenamefont {Ratner}(2006)}]{Goldsmith:JPCB:2006}%
  \BibitemOpen
  \bibfield  {author} {\bibinfo {author} {\bibfnamefont {R.~H.}\ \bibnamefont
  {Goldsmith}}, \bibinfo {author} {\bibfnamefont {M.~R.}\ \bibnamefont
  {Wasielewski}}, \ and\ \bibinfo {author} {\bibfnamefont {M.~A.}\ \bibnamefont
  {Ratner}},\ }\href {\doibase 10.1021/jp0639187} {\bibfield  {journal}
  {\bibinfo  {journal} {J. Phys. Chem. B}\ }\textbf {\bibinfo {volume} {110}},\
  \bibinfo {pages} {20258} (\bibinfo {year} {2006})}\BibitemShut {NoStop}%
\bibitem [{\citenamefont {Paulson}\ \emph {et~al.}(2005)\citenamefont
  {Paulson}, \citenamefont {Miller}, \citenamefont {Gan},\ and\ \citenamefont
  {Closs}}]{Paulson:JACS:2005}%
  \BibitemOpen
  \bibfield  {author} {\bibinfo {author} {\bibfnamefont {B.~P.}\ \bibnamefont
  {Paulson}}, \bibinfo {author} {\bibfnamefont {J.~R.}\ \bibnamefont {Miller}},
  \bibinfo {author} {\bibfnamefont {W.-X.}\ \bibnamefont {Gan}}, \ and\
  \bibinfo {author} {\bibfnamefont {G.}~\bibnamefont {Closs}},\ }\href
  {\doibase 10.1021/ja044946a} {\bibfield  {journal} {\bibinfo  {journal} {J.
  Am. Chem. Soc.}\ }\textbf {\bibinfo {volume} {127}},\ \bibinfo {pages} {4860}
  (\bibinfo {year} {2005})}\BibitemShut {NoStop}%
\bibitem [{\citenamefont {Berkelbach}, \citenamefont {Hybertsen},\ and\
  \citenamefont {Reichman}(2013)}]{Berkelbach:JCP:2013}%
  \BibitemOpen
  \bibfield  {author} {\bibinfo {author} {\bibfnamefont {T.~C.}\ \bibnamefont
  {Berkelbach}}, \bibinfo {author} {\bibfnamefont {M.~S.}\ \bibnamefont
  {Hybertsen}}, \ and\ \bibinfo {author} {\bibfnamefont {D.~R.}\ \bibnamefont
  {Reichman}},\ }\href {\doibase 10.1063/1.4794427} {\bibfield  {journal}
  {\bibinfo  {journal} {J. Chem. Phys.}\ }\textbf {\bibinfo {volume} {138}},\
  \bibinfo {pages} {114103} (\bibinfo {year} {2013})}\BibitemShut {NoStop}%
\bibitem [{\citenamefont {Guo}\ \emph {et~al.}(2019)\citenamefont {Guo},
  \citenamefont {Ma}, \citenamefont {Niu}, \citenamefont {Zhang}, \citenamefont
  {Tao}, \citenamefont {Guo}, \citenamefont {Wang},\ and\ \citenamefont
  {Xia}}]{Guo:JACS:2019}%
  \BibitemOpen
  \bibfield  {author} {\bibinfo {author} {\bibfnamefont {Y.}~\bibnamefont
  {Guo}}, \bibinfo {author} {\bibfnamefont {Z.}~\bibnamefont {Ma}}, \bibinfo
  {author} {\bibfnamefont {X.}~\bibnamefont {Niu}}, \bibinfo {author}
  {\bibfnamefont {W.}~\bibnamefont {Zhang}}, \bibinfo {author} {\bibfnamefont
  {M.}~\bibnamefont {Tao}}, \bibinfo {author} {\bibfnamefont {Q.}~\bibnamefont
  {Guo}}, \bibinfo {author} {\bibfnamefont {Z.}~\bibnamefont {Wang}}, \ and\
  \bibinfo {author} {\bibfnamefont {A.}~\bibnamefont {Xia}},\ }\href {\doibase
  10.1021/jacs.9b05723} {\bibfield  {journal} {\bibinfo  {journal} {J. Am.
  Chem. Soc.}\ }\textbf {\bibinfo {volume} {141}},\ \bibinfo {pages} {12789}
  (\bibinfo {year} {2019})}\BibitemShut {NoStop}%
\bibitem [{\citenamefont {Arlt}\ \emph {et~al.}(1993)\citenamefont {Arlt},
  \citenamefont {Schmidt}, \citenamefont {Kaiser}, \citenamefont
  {Lauterwasser}, \citenamefont {Meyer}, \citenamefont {Scheer},\ and\
  \citenamefont {Zinth}}]{Arlt:PNAS:1993}%
  \BibitemOpen
  \bibfield  {author} {\bibinfo {author} {\bibfnamefont {T.}~\bibnamefont
  {Arlt}}, \bibinfo {author} {\bibfnamefont {S.}~\bibnamefont {Schmidt}},
  \bibinfo {author} {\bibfnamefont {W.}~\bibnamefont {Kaiser}}, \bibinfo
  {author} {\bibfnamefont {C.}~\bibnamefont {Lauterwasser}}, \bibinfo {author}
  {\bibfnamefont {M.}~\bibnamefont {Meyer}}, \bibinfo {author} {\bibfnamefont
  {H.}~\bibnamefont {Scheer}}, \ and\ \bibinfo {author} {\bibfnamefont
  {W.}~\bibnamefont {Zinth}},\ }\href {\doibase 10.1073/pnas.90.24.11757}
  {\bibfield  {journal} {\bibinfo  {journal} {Proc. Nat. Acad. Sci. USA}\
  }\textbf {\bibinfo {volume} {90}},\ \bibinfo {pages} {11757} (\bibinfo {year}
  {1993})}\BibitemShut {NoStop}%
\bibitem [{\citenamefont {Bixon}\ and\ \citenamefont
  {Jortner}(1989)}]{Bixon:CPL:1989}%
  \BibitemOpen
  \bibfield  {author} {\bibinfo {author} {\bibfnamefont {M.}~\bibnamefont
  {Bixon}}\ and\ \bibinfo {author} {\bibfnamefont {J.}~\bibnamefont
  {Jortner}},\ }\href
  {http://www.sciencedirect.com/science/article/pii/S0009261489874457}
  {\bibfield  {journal} {\bibinfo  {journal} {Chem. Phys. Lett.}\ }\textbf
  {\bibinfo {volume} {159}},\ \bibinfo {pages} {17} (\bibinfo {year}
  {1989})}\BibitemShut {NoStop}%
\bibitem [{\citenamefont {Huppmann}\ \emph {et~al.}(2003)\citenamefont
  {Huppmann}, \citenamefont {Sp{\"o}rlein}, \citenamefont {Bibikova},
  \citenamefont {Oesterhelt}, \citenamefont {Wachtveitl},\ and\ \citenamefont
  {Zinth}}]{Huppmann:JPCA:2003}%
  \BibitemOpen
  \bibfield  {author} {\bibinfo {author} {\bibfnamefont {P.}~\bibnamefont
  {Huppmann}}, \bibinfo {author} {\bibfnamefont {S.}~\bibnamefont
  {Sp{\"o}rlein}}, \bibinfo {author} {\bibfnamefont {M.}~\bibnamefont
  {Bibikova}}, \bibinfo {author} {\bibfnamefont {D.}~\bibnamefont
  {Oesterhelt}}, \bibinfo {author} {\bibfnamefont {J.}~\bibnamefont
  {Wachtveitl}}, \ and\ \bibinfo {author} {\bibfnamefont {W.}~\bibnamefont
  {Zinth}},\ }\href {https://doi.org/10.1021/jp027845c} {\bibfield  {journal}
  {\bibinfo  {journal} {J. Phys. Chem. A}\ }\textbf {\bibinfo {volume} {107}},\
  \bibinfo {pages} {8302} (\bibinfo {year} {2003})}\BibitemShut {NoStop}%
\bibitem [{\citenamefont {Fingerhut}, \citenamefont {Zinth},\ and\
  \citenamefont {de~Vivie-Riedle}(2008)}]{Fingerhut:CPL:2008}%
  \BibitemOpen
  \bibfield  {author} {\bibinfo {author} {\bibfnamefont {B.~P.}\ \bibnamefont
  {Fingerhut}}, \bibinfo {author} {\bibfnamefont {W.}~\bibnamefont {Zinth}}, \
  and\ \bibinfo {author} {\bibfnamefont {R.}~\bibnamefont {de~Vivie-Riedle}},\
  }\href {http://www.sciencedirect.com/science/article/pii/S0009261408014218}
  {\bibfield  {journal} {\bibinfo  {journal} {Chem. Phys. Lett.}\ }\textbf
  {\bibinfo {volume} {466}},\ \bibinfo {pages} {209} (\bibinfo {year}
  {2008})}\BibitemShut {NoStop}%
\bibitem [{\citenamefont {Vos}\ \emph {et~al.}(1993)\citenamefont {Vos},
  \citenamefont {Rappaport}, \citenamefont {Lambry}, \citenamefont {Breton},\
  and\ \citenamefont {Martin}}]{Vos:Nature:1993}%
  \BibitemOpen
  \bibfield  {author} {\bibinfo {author} {\bibfnamefont {M.~H.}\ \bibnamefont
  {Vos}}, \bibinfo {author} {\bibfnamefont {F.}~\bibnamefont {Rappaport}},
  \bibinfo {author} {\bibfnamefont {J.-C.}\ \bibnamefont {Lambry}}, \bibinfo
  {author} {\bibfnamefont {J.}~\bibnamefont {Breton}}, \ and\ \bibinfo {author}
  {\bibfnamefont {J.-L.}\ \bibnamefont {Martin}},\ }\href {\doibase
  10.1038/363320a0} {\bibfield  {journal} {\bibinfo  {journal} {Nature}\
  }\textbf {\bibinfo {volume} {363}},\ \bibinfo {pages} {320} (\bibinfo {year}
  {1993})}\BibitemShut {NoStop}%
\bibitem [{\citenamefont {Sp{\"o}rlein}, \citenamefont {Zinth},\ and\
  \citenamefont {Wachtveitl}(1998)}]{Sporlein:JPCB:1998}%
  \BibitemOpen
  \bibfield  {author} {\bibinfo {author} {\bibfnamefont {S.}~\bibnamefont
  {Sp{\"o}rlein}}, \bibinfo {author} {\bibfnamefont {W.}~\bibnamefont {Zinth}},
  \ and\ \bibinfo {author} {\bibfnamefont {J.}~\bibnamefont {Wachtveitl}},\
  }\href {\doibase 10.1021/jp9817473} {\bibfield  {journal} {\bibinfo
  {journal} {J. Phys. Chem. B}\ }\textbf {\bibinfo {volume} {102}},\ \bibinfo
  {pages} {7492} (\bibinfo {year} {1998})}\BibitemShut {NoStop}%
\bibitem [{\citenamefont {Novoderezhkin}\ \emph {et~al.}(2004)\citenamefont
  {Novoderezhkin}, \citenamefont {Yakovlev}, \citenamefont {van Grondelle},\
  and\ \citenamefont {Shuvalov}}]{Novoderezhkin:JPCB:2004}%
  \BibitemOpen
  \bibfield  {author} {\bibinfo {author} {\bibfnamefont {V.~I.}\ \bibnamefont
  {Novoderezhkin}}, \bibinfo {author} {\bibfnamefont {A.~G.}\ \bibnamefont
  {Yakovlev}}, \bibinfo {author} {\bibfnamefont {R.}~\bibnamefont {van
  Grondelle}}, \ and\ \bibinfo {author} {\bibfnamefont {V.~A.}\ \bibnamefont
  {Shuvalov}},\ }\href {https://doi.org/10.1021/jp0373346} {\bibfield
  {journal} {\bibinfo  {journal} {J. Phys. Chem. B}\ }\textbf {\bibinfo
  {volume} {108}},\ \bibinfo {pages} {7445} (\bibinfo {year}
  {2004})}\BibitemShut {NoStop}%
\bibitem [{\citenamefont {Ma}\ \emph {et~al.}(2019)\citenamefont {Ma},
  \citenamefont {Romero}, \citenamefont {Jones}, \citenamefont
  {Novoderezhkin},\ and\ \citenamefont {van Grondelle}}]{Ma:NatCom:2019}%
  \BibitemOpen
  \bibfield  {author} {\bibinfo {author} {\bibfnamefont {F.}~\bibnamefont
  {Ma}}, \bibinfo {author} {\bibfnamefont {E.}~\bibnamefont {Romero}}, \bibinfo
  {author} {\bibfnamefont {M.~R.}\ \bibnamefont {Jones}}, \bibinfo {author}
  {\bibfnamefont {V.~I.}\ \bibnamefont {Novoderezhkin}}, \ and\ \bibinfo
  {author} {\bibfnamefont {R.}~\bibnamefont {van Grondelle}},\ }\href {\doibase
  10.1038/s41467-019-08751-8} {\bibfield  {journal} {\bibinfo  {journal} {Nat.
  Commun.}\ }\textbf {\bibinfo {volume} {10}},\ \bibinfo {pages} {933}
  (\bibinfo {year} {2019})}\BibitemShut {NoStop}%
\bibitem [{\citenamefont {Tanaka}\ and\ \citenamefont
  {Tanimura}(2009)}]{Tanaka:JPJap:2009}%
  \BibitemOpen
  \bibfield  {author} {\bibinfo {author} {\bibfnamefont {M.}~\bibnamefont
  {Tanaka}}\ and\ \bibinfo {author} {\bibfnamefont {Y.}~\bibnamefont
  {Tanimura}},\ }\href {\doibase 10.1143/JPSJ.78.073802} {\bibfield  {journal}
  {\bibinfo  {journal} {J. Phys. Soc. Jpn.}\ }\textbf {\bibinfo {volume}
  {78}},\ \bibinfo {pages} {073802} (\bibinfo {year} {2009})}\BibitemShut
  {NoStop}%
\bibitem [{\citenamefont {Tanaka}\ and\ \citenamefont
  {Tanimura}(2010)}]{Tanaka:JCP:2010}%
  \BibitemOpen
  \bibfield  {author} {\bibinfo {author} {\bibfnamefont {M.}~\bibnamefont
  {Tanaka}}\ and\ \bibinfo {author} {\bibfnamefont {Y.}~\bibnamefont
  {Tanimura}},\ }\href {\doibase 10.1063/1.3428674} {\bibfield  {journal}
  {\bibinfo  {journal} {J. Chem. Phys.}\ }\textbf {\bibinfo {volume} {132}},\
  \bibinfo {pages} {214502} (\bibinfo {year} {2010})}\BibitemShut {NoStop}%
\bibitem [{\citenamefont {Kramer}, \citenamefont {Rodr{\'\i}guez},\ and\
  \citenamefont {Zelinskyy}(2017)}]{Kramer:JPCB:2017}%
  \BibitemOpen
  \bibfield  {author} {\bibinfo {author} {\bibfnamefont {T.}~\bibnamefont
  {Kramer}}, \bibinfo {author} {\bibfnamefont {M.}~\bibnamefont
  {Rodr{\'\i}guez}}, \ and\ \bibinfo {author} {\bibfnamefont {Y.}~\bibnamefont
  {Zelinskyy}},\ }\href {\doibase 10.1021/acs.jpcb.6b09858} {\bibfield
  {journal} {\bibinfo  {journal} {J. Phys. Chem. B}\ }\textbf {\bibinfo
  {volume} {121}},\ \bibinfo {pages} {463} (\bibinfo {year}
  {2017})}\BibitemShut {NoStop}%
\bibitem [{\citenamefont {Lambert}\ \emph {et~al.}(2019)\citenamefont
  {Lambert}, \citenamefont {Ahmed}, \citenamefont {Cirio},\ and\ \citenamefont
  {Nori}}]{Lambert:NatCom:2019}%
  \BibitemOpen
  \bibfield  {author} {\bibinfo {author} {\bibfnamefont {N.}~\bibnamefont
  {Lambert}}, \bibinfo {author} {\bibfnamefont {S.}~\bibnamefont {Ahmed}},
  \bibinfo {author} {\bibfnamefont {M.}~\bibnamefont {Cirio}}, \ and\ \bibinfo
  {author} {\bibfnamefont {F.}~\bibnamefont {Nori}},\ }\href {\doibase
  10.1038/s41467-019-11656-1} {\bibfield  {journal} {\bibinfo  {journal} {Nat.
  Commun.}\ }\textbf {\bibinfo {volume} {10}},\ \bibinfo {pages} {3721}
  (\bibinfo {year} {2019})}\BibitemShut {NoStop}%
\bibitem [{\citenamefont {Makri}\ \emph {et~al.}(1996)\citenamefont {Makri},
  \citenamefont {Sim}, \citenamefont {Makarov},\ and\ \citenamefont
  {Topaler}}]{Makri:PNAS:1996}%
  \BibitemOpen
  \bibfield  {author} {\bibinfo {author} {\bibfnamefont {N.}~\bibnamefont
  {Makri}}, \bibinfo {author} {\bibfnamefont {E.}~\bibnamefont {Sim}}, \bibinfo
  {author} {\bibfnamefont {D.~E.}\ \bibnamefont {Makarov}}, \ and\ \bibinfo
  {author} {\bibfnamefont {M.}~\bibnamefont {Topaler}},\ }\href {\doibase
  10.1073/pnas.93.9.3926} {\bibfield  {journal} {\bibinfo  {journal} {Proc.
  Nat. Acad. Sci. USA}\ }\textbf {\bibinfo {volume} {93}},\ \bibinfo {pages}
  {3926} (\bibinfo {year} {1996})}\BibitemShut {NoStop}%
\bibitem [{\citenamefont {Sim}\ and\ \citenamefont
  {Makri}(1997{\natexlab{a}})}]{Sim:JPCB:1997}%
  \BibitemOpen
  \bibfield  {author} {\bibinfo {author} {\bibfnamefont {E.}~\bibnamefont
  {Sim}}\ and\ \bibinfo {author} {\bibfnamefont {N.}~\bibnamefont {Makri}},\
  }\href {\doibase 10.1021/jp970707g} {\bibfield  {journal} {\bibinfo
  {journal} {J. Phys. Chem. B}\ }\textbf {\bibinfo {volume} {101}},\ \bibinfo
  {pages} {5446} (\bibinfo {year} {1997}{\natexlab{a}})}\BibitemShut {NoStop}%
\bibitem [{\citenamefont {Richter}\ and\ \citenamefont
  {Fingerhut}(2017)}]{Richter:2017}%
  \BibitemOpen
  \bibfield  {author} {\bibinfo {author} {\bibfnamefont {M.}~\bibnamefont
  {Richter}}\ and\ \bibinfo {author} {\bibfnamefont {B.~P.}\ \bibnamefont
  {Fingerhut}},\ }\href {\doibase 10.1063/1.4984075} {\bibfield  {journal}
  {\bibinfo  {journal} {J. Chem. Phys.}\ }\textbf {\bibinfo {volume} {146}},\
  \bibinfo {pages} {214101} (\bibinfo {year} {2017})}\BibitemShut {NoStop}%
\bibitem [{\citenamefont {{Strathearn}}\ \emph {et~al.}(2018)\citenamefont
  {{Strathearn}}, \citenamefont {{Kirton}}, \citenamefont {{Kilda}},
  \citenamefont {{Keeling}},\ and\ \citenamefont {{Lovett}}}]{Strathearn:2018}%
  \BibitemOpen
  \bibfield  {author} {\bibinfo {author} {\bibfnamefont {A.}~\bibnamefont
  {{Strathearn}}}, \bibinfo {author} {\bibfnamefont {P.}~\bibnamefont
  {{Kirton}}}, \bibinfo {author} {\bibfnamefont {D.}~\bibnamefont {{Kilda}}},
  \bibinfo {author} {\bibfnamefont {J.}~\bibnamefont {{Keeling}}}, \ and\
  \bibinfo {author} {\bibfnamefont {B.~W.}\ \bibnamefont {{Lovett}}},\
  }\href@noop {} {\bibfield  {journal} {\bibinfo  {journal} {Nat. Commun.}\
  }\textbf {\bibinfo {volume} {9}},\ \bibinfo {eid} {3322} (\bibinfo {year}
  {2018})}\BibitemShut {NoStop}%
\bibitem [{\citenamefont {Makri}(2020)}]{Makri:JCP:2020}%
  \BibitemOpen
  \bibfield  {author} {\bibinfo {author} {\bibfnamefont {N.}~\bibnamefont
  {Makri}},\ }\href {\doibase 10.1063/1.5139473} {\bibfield  {journal}
  {\bibinfo  {journal} {J. Chem. Phys.}\ }\textbf {\bibinfo {volume} {152}},\
  \bibinfo {pages} {041104} (\bibinfo {year} {2020})}\BibitemShut {NoStop}%
\bibitem [{\citenamefont {Richter}\ and\ \citenamefont
  {Fingerhut}(2019)}]{Richter:FaradayDisc:2019}%
  \BibitemOpen
  \bibfield  {author} {\bibinfo {author} {\bibfnamefont {M.}~\bibnamefont
  {Richter}}\ and\ \bibinfo {author} {\bibfnamefont {B.~P.}\ \bibnamefont
  {Fingerhut}},\ }\href@noop {} {\bibfield  {journal} {\bibinfo  {journal}
  {Faraday Discuss.}\ }\textbf {\bibinfo {volume} {216}},\ \bibinfo {pages}
  {72} (\bibinfo {year} {2019})}\BibitemShut {NoStop}%
\bibitem [{\citenamefont {Goldstein}, \citenamefont {Franzen},\ and\
  \citenamefont {Bialek}(1993)}]{Goldstein:JPC:1993}%
  \BibitemOpen
  \bibfield  {author} {\bibinfo {author} {\bibfnamefont {R.~F.}\ \bibnamefont
  {Goldstein}}, \bibinfo {author} {\bibfnamefont {S.}~\bibnamefont {Franzen}},
  \ and\ \bibinfo {author} {\bibfnamefont {W.}~\bibnamefont {Bialek}},\ }\href
  {\doibase 10.1021/j100145a009} {\bibfield  {journal} {\bibinfo  {journal} {J.
  Phys. Chem.}\ }\textbf {\bibinfo {volume} {97}},\ \bibinfo {pages} {11168}
  (\bibinfo {year} {1993})}\BibitemShut {NoStop}%
\bibitem [{\citenamefont {Milischuk}\ and\ \citenamefont
  {Matyushov}(2003)}]{Milischuk:JCP:2003}%
  \BibitemOpen
  \bibfield  {author} {\bibinfo {author} {\bibfnamefont {A.}~\bibnamefont
  {Milischuk}}\ and\ \bibinfo {author} {\bibfnamefont {D.~V.}\ \bibnamefont
  {Matyushov}},\ }\href {\doibase 10.1063/1.1555635} {\bibfield  {journal}
  {\bibinfo  {journal} {J. Chem. Phys.}\ }\textbf {\bibinfo {volume} {118}},\
  \bibinfo {pages} {5596} (\bibinfo {year} {2003})}\BibitemShut {NoStop}%
\bibitem [{\citenamefont {Tamura}\ \emph {et~al.}(2012)\citenamefont {Tamura},
  \citenamefont {Martinazzo}, \citenamefont {Ruckenbauer},\ and\ \citenamefont
  {Burghardt}}]{Tamura:JCP:2012}%
  \BibitemOpen
  \bibfield  {author} {\bibinfo {author} {\bibfnamefont {H.}~\bibnamefont
  {Tamura}}, \bibinfo {author} {\bibfnamefont {R.}~\bibnamefont {Martinazzo}},
  \bibinfo {author} {\bibfnamefont {M.}~\bibnamefont {Ruckenbauer}}, \ and\
  \bibinfo {author} {\bibfnamefont {I.}~\bibnamefont {Burghardt}},\ }\href
  {\doibase 10.1063/1.4751486} {\bibfield  {journal} {\bibinfo  {journal} {J.
  Chem. Phys.}\ }\textbf {\bibinfo {volume} {137}},\ \bibinfo {pages} {22A540}
  (\bibinfo {year} {2012})}\BibitemShut {NoStop}%
\bibitem [{\citenamefont {Yao}\ \emph {et~al.}(2015)\citenamefont {Yao},
  \citenamefont {Zhou}, \citenamefont {Prior},\ and\ \citenamefont
  {Zhao}}]{Yao:2015aa}%
  \BibitemOpen
  \bibfield  {author} {\bibinfo {author} {\bibfnamefont {Y.}~\bibnamefont
  {Yao}}, \bibinfo {author} {\bibfnamefont {N.}~\bibnamefont {Zhou}}, \bibinfo
  {author} {\bibfnamefont {J.}~\bibnamefont {Prior}}, \ and\ \bibinfo {author}
  {\bibfnamefont {Y.}~\bibnamefont {Zhao}},\ }\href {\doibase
  10.1038/srep14555} {\bibfield  {journal} {\bibinfo  {journal} {Sci. Rep.}\
  }\textbf {\bibinfo {volume} {5}},\ \bibinfo {pages} {14555} (\bibinfo {year}
  {2015})}\BibitemShut {NoStop}%
\bibitem [{\citenamefont {Mavros}, \citenamefont {Hait},\ and\ \citenamefont
  {Van~Voorhis}(2016)}]{Mavros:2016}%
  \BibitemOpen
  \bibfield  {author} {\bibinfo {author} {\bibfnamefont {M.~G.}\ \bibnamefont
  {Mavros}}, \bibinfo {author} {\bibfnamefont {D.}~\bibnamefont {Hait}}, \ and\
  \bibinfo {author} {\bibfnamefont {T.}~\bibnamefont {Van~Voorhis}},\ }\href
  {\doibase 10.1063/1.4971166} {\bibfield  {journal} {\bibinfo  {journal} {J.
  Chem. Phys.}\ }\textbf {\bibinfo {volume} {145}},\ \bibinfo {pages} {214105}
  (\bibinfo {year} {2016})}\BibitemShut {NoStop}%
\bibitem [{\citenamefont {Nalbach}\ \emph {et~al.}(2012)\citenamefont
  {Nalbach}, \citenamefont {Pugliesi}, \citenamefont {Langhals},\ and\
  \citenamefont {Thorwart}}]{Nalbach:PRL:2012}%
  \BibitemOpen
  \bibfield  {author} {\bibinfo {author} {\bibfnamefont {P.}~\bibnamefont
  {Nalbach}}, \bibinfo {author} {\bibfnamefont {I.}~\bibnamefont {Pugliesi}},
  \bibinfo {author} {\bibfnamefont {H.}~\bibnamefont {Langhals}}, \ and\
  \bibinfo {author} {\bibfnamefont {M.}~\bibnamefont {Thorwart}},\ }\href@noop
  {} {\bibfield  {journal} {\bibinfo  {journal} {Phys. Rev. Lett.}\ }\textbf
  {\bibinfo {volume} {108}},\ \bibinfo {pages} {218302} (\bibinfo {year}
  {2012})}\BibitemShut {NoStop}%
\bibitem [{\citenamefont {Skourtis}, \citenamefont {Waldeck},\ and\
  \citenamefont {Beratan}(2010)}]{Skourtis:AnnRevPhysChem:2010}%
  \BibitemOpen
  \bibfield  {author} {\bibinfo {author} {\bibfnamefont {S.~S.}\ \bibnamefont
  {Skourtis}}, \bibinfo {author} {\bibfnamefont {D.~H.}\ \bibnamefont
  {Waldeck}}, \ and\ \bibinfo {author} {\bibfnamefont {D.~N.}\ \bibnamefont
  {Beratan}},\ }\href {\doibase 10.1146/annurev.physchem.012809.103436}
  {\bibfield  {journal} {\bibinfo  {journal} {Ann. Rev. Phys. Chem.}\ }\textbf
  {\bibinfo {volume} {61}},\ \bibinfo {pages} {461} (\bibinfo {year}
  {2010})}\BibitemShut {NoStop}%
\bibitem [{\citenamefont {Garg}, \citenamefont {Onuchic},\ and\ \citenamefont
  {Ambegaokar}(1985)}]{Garg:JCP:1985}%
  \BibitemOpen
  \bibfield  {author} {\bibinfo {author} {\bibfnamefont {A.}~\bibnamefont
  {Garg}}, \bibinfo {author} {\bibfnamefont {J.}~\bibnamefont {Onuchic}}, \
  and\ \bibinfo {author} {\bibfnamefont {V.}~\bibnamefont {Ambegaokar}},\
  }\href {\doibase 10.1063/1.449017} {\bibfield  {journal} {\bibinfo  {journal}
  {J. Chem. Phys.}\ }\textbf {\bibinfo {volume} {83}},\ \bibinfo {pages} {4491}
  (\bibinfo {year} {1985})}\BibitemShut {NoStop}%
\bibitem [{\citenamefont {Correa}\ \emph {et~al.}(2019)\citenamefont {Correa},
  \citenamefont {Xu}, \citenamefont {Morris},\ and\ \citenamefont
  {Adesso}}]{Correa:JCP:2019}%
  \BibitemOpen
  \bibfield  {author} {\bibinfo {author} {\bibfnamefont {L.~A.}\ \bibnamefont
  {Correa}}, \bibinfo {author} {\bibfnamefont {B.}~\bibnamefont {Xu}}, \bibinfo
  {author} {\bibfnamefont {B.}~\bibnamefont {Morris}}, \ and\ \bibinfo {author}
  {\bibfnamefont {G.}~\bibnamefont {Adesso}},\ }\href {\doibase
  10.1063/1.5114690} {\bibfield  {journal} {\bibinfo  {journal} {J. Chem.
  Phys.}\ }\textbf {\bibinfo {volume} {151}},\ \bibinfo {pages} {094107}
  (\bibinfo {year} {2019})}\BibitemShut {NoStop}%
\bibitem [{\citenamefont {Chernyak}\ and\ \citenamefont
  {Mukamel}(1996)}]{Chernyak:1996}%
  \BibitemOpen
  \bibfield  {author} {\bibinfo {author} {\bibfnamefont {V.}~\bibnamefont
  {Chernyak}}\ and\ \bibinfo {author} {\bibfnamefont {S.}~\bibnamefont
  {Mukamel}},\ }\href@noop {} {\bibfield  {journal} {\bibinfo  {journal} {J.
  Chem. Phys.}\ }\textbf {\bibinfo {volume} {105}},\ \bibinfo {pages} {4565}
  (\bibinfo {year} {1996})}\BibitemShut {NoStop}%
\bibitem [{\citenamefont {Makri}\ and\ \citenamefont
  {Makarov}(1995)}]{Makri:JCP:1995}%
  \BibitemOpen
  \bibfield  {author} {\bibinfo {author} {\bibfnamefont {N.}~\bibnamefont
  {Makri}}\ and\ \bibinfo {author} {\bibfnamefont {D.~E.}\ \bibnamefont
  {Makarov}},\ }\href {\doibase 10.1063/1.469509} {\bibfield  {journal}
  {\bibinfo  {journal} {J. Chem. Phys.}\ }\textbf {\bibinfo {volume} {102}},\
  \bibinfo {pages} {4611} (\bibinfo {year} {1995})}\BibitemShut {NoStop}%
\bibitem [{\citenamefont {Sim}\ and\ \citenamefont
  {Makri}(1997{\natexlab{b}})}]{Sim:CPC:1997}%
  \BibitemOpen
  \bibfield  {author} {\bibinfo {author} {\bibfnamefont {E.}~\bibnamefont
  {Sim}}\ and\ \bibinfo {author} {\bibfnamefont {N.}~\bibnamefont {Makri}},\
  }\href@noop {} {\bibfield  {journal} {\bibinfo  {journal} {Comput. Phys.
  Commun.}\ }\textbf {\bibinfo {volume} {99}},\ \bibinfo {pages} {335}
  (\bibinfo {year} {1997}{\natexlab{b}})}\BibitemShut {NoStop}%
\bibitem [{\citenamefont {Sim}(2001)}]{Sim:JCP:2001}%
  \BibitemOpen
  \bibfield  {author} {\bibinfo {author} {\bibfnamefont {E.}~\bibnamefont
  {Sim}},\ }\href@noop {} {\bibfield  {journal} {\bibinfo  {journal} {J. Chem.
  Phys.}\ }\textbf {\bibinfo {volume} {115}},\ \bibinfo {pages} {4450}
  (\bibinfo {year} {2001})}\BibitemShut {NoStop}%
\bibitem [{\citenamefont {L{\"u}}\ \emph {et~al.}(2013)\citenamefont {L{\"u}},
  \citenamefont {Duan}, \citenamefont {Li}, \citenamefont {Shenai},\ and\
  \citenamefont {Zhao}}]{Lu:JCP:2013}%
  \BibitemOpen
  \bibfield  {author} {\bibinfo {author} {\bibfnamefont {Z.}~\bibnamefont
  {L{\"u}}}, \bibinfo {author} {\bibfnamefont {L.}~\bibnamefont {Duan}},
  \bibinfo {author} {\bibfnamefont {X.}~\bibnamefont {Li}}, \bibinfo {author}
  {\bibfnamefont {P.~M.}\ \bibnamefont {Shenai}}, \ and\ \bibinfo {author}
  {\bibfnamefont {Y.}~\bibnamefont {Zhao}},\ }\href {\doibase
  10.1063/1.4825205} {\bibfield  {journal} {\bibinfo  {journal} {J. Chem.
  Phys.}\ }\textbf {\bibinfo {volume} {139}},\ \bibinfo {pages} {164103}
  (\bibinfo {year} {2013})}\BibitemShut {NoStop}%
\bibitem [{\citenamefont {Olbrich}\ \emph {et~al.}(2011)\citenamefont
  {Olbrich}, \citenamefont {Str{\"u}mpfer}, \citenamefont {Schulten},\ and\
  \citenamefont {Kleinekath{\"o}fer}}]{Olbrich:JPCB:2011}%
  \BibitemOpen
  \bibfield  {author} {\bibinfo {author} {\bibfnamefont {C.}~\bibnamefont
  {Olbrich}}, \bibinfo {author} {\bibfnamefont {J.}~\bibnamefont
  {Str{\"u}mpfer}}, \bibinfo {author} {\bibfnamefont {K.}~\bibnamefont
  {Schulten}}, \ and\ \bibinfo {author} {\bibfnamefont {U.}~\bibnamefont
  {Kleinekath{\"o}fer}},\ }\href@noop {} {\bibfield  {journal} {\bibinfo
  {journal} {J. Phys. Chem. B}\ }\textbf {\bibinfo {volume} {115}},\ \bibinfo
  {pages} {758} (\bibinfo {year} {2011})}\BibitemShut {NoStop}%
\bibitem [{\citenamefont {Nalbach}\ \emph {et~al.}(2017)\citenamefont
  {Nalbach}, \citenamefont {Klinkenberg}, \citenamefont {Palm},\ and\
  \citenamefont {M\"uller}}]{Nalbach:PhysRevE:2017}%
  \BibitemOpen
  \bibfield  {author} {\bibinfo {author} {\bibfnamefont {P.}~\bibnamefont
  {Nalbach}}, \bibinfo {author} {\bibfnamefont {N.}~\bibnamefont
  {Klinkenberg}}, \bibinfo {author} {\bibfnamefont {T.}~\bibnamefont {Palm}}, \
  and\ \bibinfo {author} {\bibfnamefont {N.}~\bibnamefont {M\"uller}},\
  }\href@noop {} {\bibfield  {journal} {\bibinfo  {journal} {Phys. Rev. E}\
  }\textbf {\bibinfo {volume} {96}},\ \bibinfo {pages} {042134} (\bibinfo
  {year} {2017})}\BibitemShut {NoStop}%
\bibitem [{\citenamefont {Krempl}, \citenamefont {Domcke},\ and\ \citenamefont
  {Winterstetter}(1996)}]{Krempl:ChemPhys:1996}%
  \BibitemOpen
  \bibfield  {author} {\bibinfo {author} {\bibfnamefont {S.}~\bibnamefont
  {Krempl}}, \bibinfo {author} {\bibfnamefont {W.}~\bibnamefont {Domcke}}, \
  and\ \bibinfo {author} {\bibfnamefont {M.}~\bibnamefont {Winterstetter}},\
  }\href {\doibase https://doi.org/10.1016/0301-0104(96)00016-X} {\bibfield
  {journal} {\bibinfo  {journal} {Chem. Phys.}\ }\textbf {\bibinfo {volume}
  {206}},\ \bibinfo {pages} {63} (\bibinfo {year} {1996})}\BibitemShut
  {NoStop}%
\bibitem [{\citenamefont {Palm}\ and\ \citenamefont
  {Nalbach}(2018)}]{Palm:JCP:2018}%
  \BibitemOpen
  \bibfield  {author} {\bibinfo {author} {\bibfnamefont {T.}~\bibnamefont
  {Palm}}\ and\ \bibinfo {author} {\bibfnamefont {P.}~\bibnamefont {Nalbach}},\
  }\href {\doibase 10.1063/1.5051652} {\bibfield  {journal} {\bibinfo
  {journal} {J. Chem. Phys.}\ }\textbf {\bibinfo {volume} {149}},\ \bibinfo
  {pages} {214103} (\bibinfo {year} {2018})}\BibitemShut {NoStop}%
\bibitem [{\citenamefont {Domcke}, \citenamefont {Yarkony},\ and\ \citenamefont
  {K{\"o}ppel}(2004)}]{Domcke:Book}%
  \BibitemOpen
  \bibfield  {author} {\bibinfo {author} {\bibfnamefont {W.}~\bibnamefont
  {Domcke}}, \bibinfo {author} {\bibfnamefont {D.~R.}\ \bibnamefont {Yarkony}},
  \ and\ \bibinfo {author} {\bibfnamefont {H.}~\bibnamefont {K{\"o}ppel}},\
  }\href@noop {} {\emph {\bibinfo {title} {Conical Intersections}}},\ Advanced
  Series in Physical Chemistry\ (\bibinfo  {publisher} {World Scientific},\
  \bibinfo {year} {2004})\BibitemShut {NoStop}%
\bibitem [{\citenamefont {Romero-Rochin}\ and\ \citenamefont
  {Oppenheim}(1989)}]{Romero-Rochin:PhysicaA:1989}%
  \BibitemOpen
  \bibfield  {author} {\bibinfo {author} {\bibfnamefont {V.}~\bibnamefont
  {Romero-Rochin}}\ and\ \bibinfo {author} {\bibfnamefont {I.}~\bibnamefont
  {Oppenheim}},\ }\href@noop {} {\bibfield  {journal} {\bibinfo  {journal}
  {Physica A}\ }\textbf {\bibinfo {volume} {155}},\ \bibinfo {pages} {52}
  (\bibinfo {year} {1989})}\BibitemShut {NoStop}%
\bibitem [{\citenamefont {Fingerhut}\ and\ \citenamefont
  {Mukamel}(2012)}]{Fingerhut:JPCL:2012}%
  \BibitemOpen
  \bibfield  {author} {\bibinfo {author} {\bibfnamefont {B.~P.}\ \bibnamefont
  {Fingerhut}}\ and\ \bibinfo {author} {\bibfnamefont {S.}~\bibnamefont
  {Mukamel}},\ }\href {\doibase 10.1021/jz3006282} {\bibfield  {journal}
  {\bibinfo  {journal} {J. Phys. Chem. Lett.}\ }\textbf {\bibinfo {volume}
  {3}},\ \bibinfo {pages} {1798} (\bibinfo {year} {2012})}\BibitemShut
  {NoStop}%
\bibitem [{\citenamefont {Lambert}\ and\ \citenamefont
  {Makri}(2012)}]{Lambert:MolPhys:2012}%
  \BibitemOpen
  \bibfield  {author} {\bibinfo {author} {\bibfnamefont {R.}~\bibnamefont
  {Lambert}}\ and\ \bibinfo {author} {\bibfnamefont {N.}~\bibnamefont
  {Makri}},\ }\href@noop {} {\bibfield  {journal} {\bibinfo  {journal} {Mol.
  Phys.}\ }\textbf {\bibinfo {volume} {110}},\ \bibinfo {pages} {1967}
  (\bibinfo {year} {2012})}\BibitemShut {NoStop}%
\bibitem [{\citenamefont {Bixon}, \citenamefont {Jortner},\ and\ \citenamefont
  {Michel-Beyerle}(1995)}]{Bixon:ChemPhys:1995}%
  \BibitemOpen
  \bibfield  {author} {\bibinfo {author} {\bibfnamefont {M.}~\bibnamefont
  {Bixon}}, \bibinfo {author} {\bibfnamefont {J.}~\bibnamefont {Jortner}}, \
  and\ \bibinfo {author} {\bibfnamefont {M.~E.}\ \bibnamefont
  {Michel-Beyerle}},\ }\href {\doibase
  https://doi.org/10.1016/0301-0104(95)00168-N} {\bibfield  {journal} {\bibinfo
   {journal} {Chem. Phys.}\ }\textbf {\bibinfo {volume} {197}},\ \bibinfo
  {pages} {389} (\bibinfo {year} {1995})}\BibitemShut {NoStop}%
\bibitem [{Note1()}]{Note1}%
  \BibitemOpen
  \bibinfo {note} {We adopt the definition of Ref.~\protect \citenum
  {Bixon:JCP:1997} where a superexchange mechanism is reflected in a vanishing
  transient population of the bridge $B$. This definition differs from
  Ref.~\protect \citenum {Skourtis:ChemPhys:1995} where superexchange mechanism
  is mediated by the off-diagonal elements of the density matrix.}\BibitemShut
  {Stop}%
\bibitem [{\citenamefont {Weiss}(2012)}]{Weiss:book}%
  \BibitemOpen
  \bibfield  {author} {\bibinfo {author} {\bibfnamefont {U.}~\bibnamefont
  {Weiss}},\ }\href@noop {} {\emph {\bibinfo {title} {{Quantum Dissipative
  Systems; 4th ed.}}}}\ (\bibinfo  {publisher} {World Scientific},\ \bibinfo
  {address} {Singapore},\ \bibinfo {year} {2012})\BibitemShut {NoStop}%
\bibitem [{\citenamefont {Chakravarty}(1982)}]{Chakravarty:1982}%
  \BibitemOpen
  \bibfield  {author} {\bibinfo {author} {\bibfnamefont {S.}~\bibnamefont
  {Chakravarty}},\ }\href {\doibase 10.1103/PhysRevLett.49.681} {\bibfield
  {journal} {\bibinfo  {journal} {Phys. Rev. Lett.}\ }\textbf {\bibinfo
  {volume} {49}},\ \bibinfo {pages} {681} (\bibinfo {year} {1982})}\BibitemShut
  {NoStop}%
\bibitem [{\citenamefont {Bray}\ and\ \citenamefont {Moore}(1982)}]{Bray:1982}%
  \BibitemOpen
  \bibfield  {author} {\bibinfo {author} {\bibfnamefont {A.~J.}\ \bibnamefont
  {Bray}}\ and\ \bibinfo {author} {\bibfnamefont {M.~A.}\ \bibnamefont
  {Moore}},\ }\href {\doibase 10.1103/PhysRevLett.49.1545} {\bibfield
  {journal} {\bibinfo  {journal} {Phys. Rev. Lett.}\ }\textbf {\bibinfo
  {volume} {49}},\ \bibinfo {pages} {1545} (\bibinfo {year}
  {1982})}\BibitemShut {NoStop}%
\bibitem [{\citenamefont {Acharyya}, \citenamefont {Richter},\ and\
  \citenamefont {Fingerhut}(2020)}]{Nirmal:2020}%
  \BibitemOpen
  \bibfield  {author} {\bibinfo {author} {\bibfnamefont {N.}~\bibnamefont
  {Acharyya}}, \bibinfo {author} {\bibfnamefont {M.}~\bibnamefont {Richter}}, \
  and\ \bibinfo {author} {\bibfnamefont {B.~P.}\ \bibnamefont {Fingerhut}},\
  }\href@noop {} {\bibfield  {journal} {\bibinfo  {journal} {arXiv:2009.12296}\
  } (\bibinfo {year} {2020})}\BibitemShut {NoStop}%
\bibitem [{\citenamefont {Aghtar}\ \emph {et~al.}(2017)\citenamefont {Aghtar},
  \citenamefont {Kleinekath{\"o}fer}, \citenamefont {Curutchet},\ and\
  \citenamefont {Mennucci}}]{Aghtar:JPCB:2017}%
  \BibitemOpen
  \bibfield  {author} {\bibinfo {author} {\bibfnamefont {M.}~\bibnamefont
  {Aghtar}}, \bibinfo {author} {\bibfnamefont {U.}~\bibnamefont
  {Kleinekath{\"o}fer}}, \bibinfo {author} {\bibfnamefont {C.}~\bibnamefont
  {Curutchet}}, \ and\ \bibinfo {author} {\bibfnamefont {B.}~\bibnamefont
  {Mennucci}},\ }\href@noop {} {\bibfield  {journal} {\bibinfo  {journal} {J.
  Phys. Chem. B}\ }\textbf {\bibinfo {volume} {121}},\ \bibinfo {pages} {1330}
  (\bibinfo {year} {2017})}\BibitemShut {NoStop}%
\bibitem [{\citenamefont {Liebel}\ \emph {et~al.}(2015)\citenamefont {Liebel},
  \citenamefont {Schnedermann}, \citenamefont {Wende},\ and\ \citenamefont
  {Kukura}}]{Liebel:JPCA:2015}%
  \BibitemOpen
  \bibfield  {author} {\bibinfo {author} {\bibfnamefont {M.}~\bibnamefont
  {Liebel}}, \bibinfo {author} {\bibfnamefont {C.}~\bibnamefont
  {Schnedermann}}, \bibinfo {author} {\bibfnamefont {T.}~\bibnamefont {Wende}},
  \ and\ \bibinfo {author} {\bibfnamefont {P.}~\bibnamefont {Kukura}},\ }\href
  {\doibase 10.1021/acs.jpca.5b05948} {\bibfield  {journal} {\bibinfo
  {journal} {J. Phys. Chem. A}\ }\textbf {\bibinfo {volume} {119}},\ \bibinfo
  {pages} {9506} (\bibinfo {year} {2015})}\BibitemShut {NoStop}%
\bibitem [{\citenamefont {Kowalewski}\ \emph {et~al.}(2017)\citenamefont
  {Kowalewski}, \citenamefont {Fingerhut}, \citenamefont {Dorfman},
  \citenamefont {Bennett},\ and\ \citenamefont
  {Mukamel}}]{Kowalewski:ChemRev:2017}%
  \BibitemOpen
  \bibfield  {author} {\bibinfo {author} {\bibfnamefont {M.}~\bibnamefont
  {Kowalewski}}, \bibinfo {author} {\bibfnamefont {B.~P.}\ \bibnamefont
  {Fingerhut}}, \bibinfo {author} {\bibfnamefont {K.~E.}\ \bibnamefont
  {Dorfman}}, \bibinfo {author} {\bibfnamefont {K.}~\bibnamefont {Bennett}}, \
  and\ \bibinfo {author} {\bibfnamefont {S.}~\bibnamefont {Mukamel}},\ }\href
  {\doibase 10.1021/acs.chemrev.7b00081} {\bibfield  {journal} {\bibinfo
  {journal} {Chem. Rev.}\ }\textbf {\bibinfo {volume} {117}},\ \bibinfo {pages}
  {12165} (\bibinfo {year} {2017})}\BibitemShut {NoStop}%
\bibitem [{\citenamefont {Skourtis}\ and\ \citenamefont
  {Mukamel}(1995)}]{Skourtis:ChemPhys:1995}%
  \BibitemOpen
  \bibfield  {author} {\bibinfo {author} {\bibfnamefont {S.~S.}\ \bibnamefont
  {Skourtis}}\ and\ \bibinfo {author} {\bibfnamefont {S.}~\bibnamefont
  {Mukamel}},\ }\href {\doibase https://doi.org/10.1016/0301-0104(95)00167-M}
  {\bibfield  {journal} {\bibinfo  {journal} {Chem. Phys.}\ }\textbf {\bibinfo
  {volume} {197}},\ \bibinfo {pages} {367} (\bibinfo {year}
  {1995})}\BibitemShut {NoStop}%
\end{thebibliography}%

\end{document}